\documentclass[aps,preprint,superscriptaddress]{revtex4-1}

\usepackage{amsmath}
\usepackage{fixmath}
\usepackage{accents}
\usepackage{graphicx}%

\usepackage{hyperref}

\allowdisplaybreaks

\newcommand{\mb}{\mathbold}

\begin{document}
\title{Second-order nonlinear optical response of graphene}
\author{Yongrui Wang}
\affiliation{Department of Physics and Astronomy, Texas A\&M
University, College Station, TX, 77843 USA}
\author{Mikhail Tokman}
\affiliation{Institute of Applied Physics, Russian Academy of Sciences}
\author{Alexey Belyanin}
\affiliation{Department of Physics and Astronomy, Texas A\&M
University, College Station, TX, 77843 USA}

\date{\today}

\begin{abstract}

Although massless Dirac fermions in graphene constitute a centrosymmetric medium for in-plane excitations, their second-order nonlinear optical response is nonzero if the effects of spatial dispersion are taken into account. Here we present a rigorous quantum-mechanical theory of the second-order nonlinear response of graphene beyond the electric dipole approximation, which includes both intraband and interband transitions. The resulting nonlinear susceptibility tensor satisfies all symmetry and permutation properties, and can be applied to all three-wave mixing processes. We obtain useful analytic expressions in the limit of a degenerate electron distribution, which reveal quite strong second-order nonlinearity at long wavelengths, Fermi-edge resonances, and unusual polarization properties. 
\end{abstract}


\maketitle

\section{Introduction}

Nonlinear optical properties of graphene have attracted considerable interest in the community. The magnitude of the matrix element of the interaction Hamiltonian describing coupling of massless Dirac electrons to light scales in proportion to $ v_F/\omega \propto \lambda$, i.e.~it grows more rapidly with wavelength $\lambda$ than in conventional materials with parabolic energy dispersion, where it scales roughly as $\sqrt{\lambda}$. This promises a strong nonlinear response at  long wavelengths.  Unfortunately, graphene is also a centrosymmetric medium for low-energy in-plane excitations, which suppresses second-order nonlinear response in the electric dipole approximation. Therefore, most of the effort was concentrated on the third-order nonlinear processes that are electric dipole-allowed. Recent theoretical proposals and some experiments include third-harmonic generation \cite{kumar2013,hong2013}, four-wave mixing \cite{hendry2010,gu2012,sun2010} and current-induced second-harmonic generation \cite{glazov2014,bykov2012,cheng2014}. In few-layer graphene, second-harmonic generation (SHG) arising from the interactions between layers, which breaks the inversion symmetry, has been observed \cite{dean2009,dean2010}. 

The aim of this paper is to show that monolayer graphene does demonstrate quite significant second-order nonlinearity at long wavelengths despite its inversion symmetry. Here and throughout the paper, we will discuss only the 2D (surface) nonlinearity due to in-plane motion of electrons. Like any surface, graphene exhibits anisotropy between in-plane and out-of-plane electron motion. However, the corresponding second-order nonlinearity is very small and we will not discuss it here. 

We develop the full quantum-mechanical theory of the in-plane second-order nonlinear response beyond the electric dipole approximation. In this case one has to consider oblique or in-plane propagation of electromagnetic waves. A non-zero in-plane second-order susceptibility $\chi^{(2)}$ of monolayer graphene appears when one includes the dependence of $\chi^{(2)}$ on the in-plane photon wave vectors, i.e. the {\it spatial dispersion}. Physically, this means that the inversion symmetry of graphene is broken by the wave vector direction. The spatial dispersion in momentum space is of course equivalent to the nonlocal response in real space. Spatial dispersion effects turn out to be quite large because of a large magnitude of the electron velocity $v_F$. A non-zero value of the nonlocal $\chi^{(2)}$ has been pointed out before for second-harmonic generation \cite{mikhailovSHG,glazov2011,smirnova2014} (which only included intraband transitions in a free-carrier model), difference-frequency generation \cite{yao2014}, and parametric frequency down-conversion \cite{tokman2016}.  The latter two papers developed a quantum theory including both intraband and interband transitions and applied it to the nonlinear generation of surface plasmons. In the recent experiment \cite{constant2016}, evidence for the difference-frequency generation of surface plasmons in graphene was reported. Here we provide a systematic derivation of the second-order nonlinear conductivity tensor, valid for all second-order processes, all frequencies and doping densities, as long as the massless Dirac fermion approximation for a single-particle Hamiltonian is applicable. For graphene, this means the range of frequencies from zero (more precisely, from inverse scattering time) to the near-infrared. Our approach can be applied to any system of massless chiral Dirac fermions, for example surface states in topological insulators such as Bi$_2$Se$_3$. The resulting nonlinear susceptibility tensor satisfies all symmetry and permutation properties, and predicts unusual polarization properties of the nonlinear signal. We also summarize main properties of the linear current as a necessary step in deriving the nonlinear response functions, and present a detailed discussion of its gauge properties and regularization. 

\section{Basic equations}

Consider a 2D quantum system which in the absence of external fields can be described by the Dirac Hamiltonian
\begin{equation}
\label{Eq:Hamiltonian}
\hat{H}_0(\hat{\mb{p}}) = v_F \hat{\mb{\sigma}} \cdot \hat{\mb{p}} ,
\end{equation}
where $\hat{\mb{p}} = \mb{x}_0 \hat{p}_x + \mb{y}_0 \hat{p}_y$, $\hat{p}_{x,y} = - i\hbar \frac{\partial}{\partial x, \partial y}$, $\hat{\bm{\sigma}} = \mb{x}_0 \hat{\sigma}_x + \mb{y}_0 \hat{\sigma}_y$, where $\hat{\sigma}_{x,y}$ are Pauli matrices. The spinor eigenfunctions $\mb{\Psi} = \begin{pmatrix} \Psi_1 \\ \Psi_2 \end{pmatrix}$ of the Hamiltonian (\ref{Eq:Hamiltonian}) are
\begin{equation}
\label{Eq:eigen_state}
\mb{\Psi}_{\mb{k},s}(\mb{r}) \equiv \langle \mb{r} |\mb{k},s \rangle = \frac{e^{i\mb{k}\cdot\mb{r}}}{\sqrt{2 A}} 
\begin{pmatrix}
s \\ e^{i\theta(\mb{k})}
\end{pmatrix} ,
\end{equation}
and the eigenenergies are $E = s \hbar v_F k$ where $s = \pm 1$ for conduction and valence bands, respectively; $\mb{k} = \mb{x}_0 k_x + \mb{y}_0 k_y$, $\theta(\mb{k})$ is the angle between the electron momentum $\mb{k}$ and $x$-axis, and $A$ is the normalization area. This description is valid for carriers in monolayer graphene up to the energies of order 1 eV. For higher energies, quadratic and trigonal warping corrections become non-negligible.

Consider the most general light-matter interaction Hamiltonian utilizing both vector and scalar potentials: $\mb{E} = -\nabla \varphi - c^{-1} \dot{\mb{A}}$ and $\mb{B} = \nabla \times \mb{A}$. Following a standard procedure \cite{landau2013quantum, gantmakher1987}, we replace $\hat{\mb{p}} \Rightarrow \hat{\mb{p}} + \frac{e}{c} \mb{A}$ in the unperturbed Hamiltonian $\hat{H}_0(\hat{\mb{p}})$ and add the potential energy operator $-e \varphi$ assuming a particle with the charge $-e$. This gives
\begin{equation}
\label{Eq:Hamiltonian_with_int}
\hat{H} = \hat{H}_0 + \hat{H}_{int}^{opt} , 
\hspace{1cm} 
\hat{H}_{int}^{opt} = \frac{e v_F}{c} \hat{\mb{\sigma}} \cdot \mb{A} - e \varphi \cdot \hat{1} ,
\end{equation}
where $\hat{1}$ is a unit 2$\times$2 matrix. The Hamiltonian in Eq.~(\ref{Eq:Hamiltonian_with_int}) leads to the von Neumann equation for the density matrix:
\begin{align}
\label{Eq:density_matrix_eq}
i\hbar \frac{\partial}{\partial t} \rho_{mn} 
= 
(E_m - E_n) \rho_{mn} 
+ \sum_{l} \left[ \left( \hat{H}_{int}^{opt} \right)_{ml} \rho_{ln} - \rho_{ml} \left( \hat{H}_{int}^{opt} \right)_{ln}  \right] ,
\end{align} 
where $|n\rangle = |\mb{k},s\rangle$.

We will consider a monochromatic electromagnetic field in plane of graphene,
\begin{equation}
\label{Eq:optical_field}
\mb{E} = \frac{1}{2} \left[ \mb{x}_0 E_x(\omega) + \mb{y}_0 E_y(\omega) \right] e^{-i\omega t + i q x} + \mathrm{C.C.}
\end{equation}
or its bichromatic combinations. The field component $\mb{z}_0 E_z$ can be ignored because neither this field component itself nor the magnetic field it generates can affect the 2D carrier motion. Furthermore, the component of the vector potential $\mb{z}_0 A_z$ which generates the z-component of the electric field $\mb{z}_0 E_z$ does not enter the Hamiltonian (\ref{Eq:Hamiltonian_with_int}) because $\hat{\mb{\sigma}} \cdot \mb{z}_0 = 0$. The field described by Eq.~(\ref{Eq:optical_field}) corresponds to the electromagnetic potentials
\begin{align}
\label{Eq:EM_potentials}
\varphi &= \frac{1}{2} \phi(\omega) e^{-i\omega t + i q x} + \mathrm{C.C.}, \nonumber \\
\mb{A} &= \frac{1}{2} \left[ \mb{x}_0 A_x(\omega) + \mb{y}_0 A_y(\omega) \right] e^{-i\omega t + i q x} + \mathrm{C.C.} .
\end{align}
Note that the P-polarized radiation can be defined through both the scalar potential,
\begin{equation}
\label{Eq:scalar_P}
\varphi = \frac{1}{2} \frac{i E_x(\omega)}{q} e^{-i\omega t + i q x} + \mathrm{C.C.} ,
\end{equation}
and the vector potential:
\begin{equation}
\label{Eq:vector_P}
\mb{A}_P = \frac{1}{2} \mb{x}_0 \frac{c E_x(\omega)}{i\omega} e^{-i\omega t + i q x} + \mathrm{C.C.} 
\end{equation}
At the same time, the S-polarized radiation, can be defined only through the vector potential:
\begin{equation}
\label{Eq:vector_S}
\mb{A}_S = \frac{1}{2} \mb{y}_0 \frac{c E_y(\omega)}{i\omega} e^{-i\omega t + i q x} + \mathrm{C.C.} 
\end{equation}

It is convenient to represent the surface current density generated in response to a harmonic field as a sum over spatial harmonics: $\mb{j}(\mb{r}) = 2^{-1} \sum_{\mb{q}} \mb{j}^{(q)} e^{i\mb{q}\cdot\mb{r}} + \mathrm{C.C.}$, where $2^{-1} \mb{j}^{(q)} = S^{-1} \int_S \mb{j}(\mb{r}) e^{-i\mb{q}\cdot\mb{r}} d^2\mb{r}$; the set of in-plane photon wave vectors $\mb{q}$ is specified by appropriate conditions on the boundary of a large area $S \gg A$. It is also convenient to choose the area $S$ to be a multiple of the normalization area $A$, so that
\begin{equation}
(2 A)^{-1/2} \int_A \mb{\Psi}_n^\ast(\mb{r}) \mb{\Psi}_m(\mb{r}) d^2 r = (2 S)^{-1/2} \int_S \mb{\Psi}_n^\ast(\mb{r}) \mb{\Psi}_m(\mb{r}) d^2 r .
\end{equation}
After calculating the matrix elements $\mb{j}_{nm}^{(q)}$ of the current density operator and solving independently the master equations (\ref{Eq:density_matrix_eq}), one can calculate the average amplitude of a given current density harmonic, which could be used as a source in Maxwell's equations or to determine the conductivity tensor:
\begin{equation}
\mb{j}^{(q)} = \sum_{mn} \mb{j}_{nm}^{(q)} \rho_{mn} .
\end{equation}

In order to evaluate $\mb{j}_{nm}^{(q)}$ we determine the velocity operator $\hat{\mb{v}} = i\hbar^{-1} \left[ \hat{H},\hat{\mb{r}} \right]$ and define the current density operator as $\hat{\mb{j}} = -e \hat{\mb{v}}$:
\begin{equation}
\hat{\mb{j}} = -e v_F \hat{\mb{\sigma}} .
\end{equation}
Next, we take into account a standard expression for the current density operator in a second-quantized formalism \cite{landau2013quantum}: $\hat{\mb{j}}(\mb{r}) = \hat{\mb{\Psi}}^\dagger \cdot \hat{\mb{j}} \cdot \hat{\mb{\Psi}}$, where $\hat{\mb{\Psi}} = \sum_n \hat{a}_n \mb{\Psi}_n(\mb{r})$ and $\hat{\mb{\Psi}}^\dagger = \sum_m \hat{a}_m^\dagger \mb{\Psi}_m^\dagger(\mb{r})$ are second-quantized operators, and $\hat{a}_m^\dagger$ and $\hat{a}_n$ are fermion creation and annihilation operators. Treating $\hat{a}_m^\dagger$ and $\hat{a}_n$ as Heisenberg operators and using $\mb{j}(\mb{r}) = \langle \hat{\mb{j}}(\mb{r}) \rangle$, $\langle \hat{a}_m^\dagger(t) \hat{a}_n(t) \rangle = \rho_{mn}(t)$, we arrive at $2^{-1} \mb{j}^{(q)} = \sum_{mn} \left(e^{-i\mb{q}\cdot\mb{r}} \hat{\mb{j}} \right)_{nm} \rho_{mn}$, which gives
\begin{equation}
2^{-1} \mb{j}_{nm}^{(q)} = \langle n | e^{-i\mb{q}\cdot\mb{r}} \hat{\mb{j}} | m \rangle .
\label{Eq:jq_matrix}
\end{equation}
To calculate the matrix elements $\mb{j}_{mn}^{(q)}$ and $\left( \hat{H}_{int}^{opt} \right)_{mn}$ we will need the following useful relationships:
\begin{align}
\left(e^{iqx}\right)_{mn} &= \frac{1}{2} \left(s_m s_n + e^{i(\theta_n-\theta_m)}\right) \delta_{\mb{k}_m,\mb{k}_n+\mb{q}} ,
\label{Eq:eiqx_matrix}
 \\
\left(\hat{\mb{\sigma}} e^{i q x}\right)_{mn} &= \frac{1}{2} \left[ (\mb{x}_0 - i\mb{y}_0) s_m e^{i\theta_n} + (\mb{x}_0 + i\mb{y}_0) s_n e^{-i\theta_m} \right] \delta_{\mb{k}_m,\mb{k}_n+\mb{q}} .
\label{Eq:sigma_eiqx_matrix}
\end{align}
The above general equations should allow one to calculate the conductivity in any order with respect to the external optical field. There is however a complication related to the fact that the model described by the effective Hamiltonian Eq.~(\ref{Eq:Hamiltonian}) contains a ''bottomless'' valence band with electrons occupying all states to $k \rightarrow \infty$. Therefore, only the converging integrals make sense:
\begin{equation}
\label{Eq:sum_to_integral}
\sum_{mn} \mb{j}_{nm}^{(q)} \rho_{mn} \Rightarrow g \sum_{s s'} \underaccent{\infty'}{\int} \frac{d^2 k'}{4\pi^2} \underaccent{\infty}{\int} \frac{d^2 k}{4\pi^2} \mb{j}^{(q)}_{\mb{k}'\mb{k}s's} \rho_{\mb{k}\mb{k}'ss'} ,
\end{equation}
where $g$ is the degeneracy factor. Otherwise the optical response could be determined by the electron dispersion far from the Dirac point where the effective Hamiltonian Eq.~(\ref{Eq:Hamiltonian}) is no longer valid. It turns out that the convergence of the linear current depends on the choice of the gauge, whereas for the second-order nonlinear current the integral in Eq.~(\ref{Eq:sum_to_integral}) converges for any gauge. The divergence of the linear response can be regularized as discussed in the next section. In addition, the gauge dependence of the linear response violates gauge invariance, which is a consequence of the fact that the density matrix corresponding to the bottomless Hamiltonian in Eq.~(\ref{Eq:Hamiltonian}) has an infinite trace. In the next section we discuss this issue in more detail.

\section{The linear response of massless Dirac fermions \label{Sec:linear}}

The perturbation expansion of the nonlinear response functions implies that the second-order nonlinear terms depend on the first-order linear response. Therefore, in this section we outline the derivation of the linear current. The nontrivial aspect of this derivation is an apparent violation of gauge invariance and divergence of the linear current. We address these issues in this section and related Appendix sections.   

The solution of the density matrix equation (\ref{Eq:density_matrix_eq}) in the linear approximation with respect to the field is
\begin{equation}
\label{Eq:rho_linear}
\rho^{(1)}_{nm}(\omega)_= \frac{1}{2} \frac{\left[ \hat{V}(\omega) e^{i q x} \right]_{nm} (n_m - n_n)}{\hbar\omega - (E_n-E_m)} ,
\end{equation}
where we defined $\hat{H}_{int}^{opt} = 2^{-1} \left[ \hat{V}(\omega) e^{-i\omega t + i q x} + \mathrm{H.C.} \right]$. \\ 
Here $\hat{V}(\omega) = -e\phi(\omega) \cdot \hat{1} + \frac{e v_F}{c} \left[ \hat{\sigma}_x A_x(\omega) + \hat{\sigma}_y A_y(\omega) \right]$ and $\rho^{(1)}_{nm}(\omega)$ is a complex-valued amplitude of the linear perturbation $\propto e^{-i\omega t}$ of the density matrix. For a monochromatic current $\mb{j} = 2^{-1} \mb{j}^{(q)}(\omega) e^{-i\omega t + i q x} + \mathrm{C.C.}$ we have
\begin{equation}
\label{Eq:jq_linear}
\mb{j}^{(q)}(\omega) = \sum_{mn} \mb{j}^{(q)}_{mn} \rho^{(1)}_{nm}(\omega) .
\end{equation}
The expression (\ref{Eq:jq_linear}) is evaluated in Appendix \ref{appendix:linear}. The most straightforward derivation is for a P-polarized field defined through a scalar potential, Eq.~(\ref{Eq:scalar_P}), since in this case the integral (\ref{Eq:sum_to_integral}) converges. If we keep only the terms of the lowest order in $q$ (i.e. the linear terms since $E_x = -i q \phi$), the resulting 2D (surface) conductivity tensor is independent of $q$. In the limit of strong degeneracy or low temperatures, the relevant terms are \\
({\romannumeral 1}) intraband conductivity, which has a Drude-like form:
\begin{equation}
\label{Eq:conductivity_linear_intra}
\sigma^{(intra)}_{xx}(\omega) = \frac{i e^2 v_F k_F}{\pi\hbar(\omega+i\gamma)} ,
\end{equation}
({\romannumeral 2}) and the interband term:
\begin{align}
\label{Eq:conductivity_linear_inter}
\sigma_{xx}^{(inter)}(\omega) = \frac{i e^2}{4\pi\hbar} \ln \left[ \frac{2 v_F k_F - (\omega + i\gamma)}{2 v_F k_F + (\omega + i\gamma)} \right] .
\end{align}
Here $k = k_F$ is Fermi momentum, and we also added the relaxation terms by replacing $\omega \rightarrow \omega + i\gamma$ in Eq.~(\ref{Eq:rho_linear}); in the limit $\gamma \rightarrow +0$ one can obtain from Eq.~(\ref{Eq:conductivity_linear_inter}) the well known result for the interband conductivity \cite{nair2008}: $\mathrm{Re} \sigma^{(inter)}_{xx} = \frac{e^2}{4\hbar} \Theta(\omega - 2 v_F k_F)$, where $\Theta(x)$ is the Heaviside step function.

If we define the optical field with a vector potential, the same calculation will lead to divergent integrals. In this case the finite, and at the same time gauge-invariant, expression for the linear current at frequency $\omega$ can be obtained by subtracting the same current evaluated at zero frequency  \cite{falkovsky2007}:
\begin{equation}
\label{Eq:jq_Falkovsky}
\mb{j}^{(q)}(\omega) = \sum_{mn} \mb{j}^{(q)}_{mn} \left[ \rho^{1,A}_{nm}(\omega) - \rho^{1,A}_{nm}(\omega \rightarrow 0) \right] .
\end{equation}
Here $\rho^{1,A}_{nm}(\omega)$ is Eq.~(\ref{Eq:rho_linear}) with $\phi(\omega) = 0$ in the interaction Hamiltonian. This prescription cancels the divergent term and leads to the Kubo formula for the linear response. In our case Eq.~(\ref{Eq:jq_Falkovsky}) is equal to the sum of Eqs.~(\ref{Eq:conductivity_linear_intra}) and (\ref{Eq:conductivity_linear_inter}) for the diagonal conductivities $\sigma_{yy} = \sigma_{xx}$, and gives $\sigma_{xy} = 0$. The procedure in Eq.~(\ref{Eq:jq_Falkovsky}) can be justified by considering the graphene Hamiltonian with a small quadratic term in the energy dispersion:
\begin{equation}
\label{Eq:dispersion_with_quad}
E = s\hbar v_F k + \epsilon \frac{\hbar^2 k^2}{2} ,
\end{equation}
where $\epsilon$ is a small parameter. Adding this term provides a bottom to the valence band. As shown in Appendix \ref{appendix:current_define}, the linear current for such a system approaches Eq.~(\ref{Eq:jq_Falkovsky}) when $\epsilon \rightarrow 0$.

For a P-polarized field which can be represented through both scalar and vector potentials the renormalization procedure in Eq.~(\ref{Eq:jq_Falkovsky}) is equivalent to the gauge transformation of the density matrix from the  A-gauge (\ref{Eq:vector_P}) to the $\varphi$-gauge (\ref{Eq:scalar_P}). Indeed, let the function $\rho^{1,A_P}_{nm}(\omega)$ correspond to the solution of Eq.~(\ref{Eq:rho_linear}) for the field defined in the gauge given by Eq.~(\ref{Eq:vector_P}), whereas the function $\rho^{1,\varphi}_{nm}(\omega)$ correspond to the gauge of Eq.~(\ref{Eq:scalar_P}). Since we just found that the sum $\sum_{mn} \mb{j}^{(q)}_{mn} \rho^{(1,\varphi)}_{nm}(\omega)$ is finite, it makes sense to try the transformation $\rho^{1,A_P}_{nm} \Rightarrow \rho^{1,\varphi}_{nm}$. The gauge transformation from $\mb{A}$ and $\varphi$ to $\tilde{\mb{A}}$ and $\tilde{\varphi}$ corresponds to the unitary transformation of the density matrix (see Appendix \ref{appendix:gauge_transform})
\begin{equation}
\label{Eq:rho_gauge_transformation}
\tilde{\rho}_{nm} = \sum_{qp} \left(e^{-\frac{ief}{\hbar c}}\right)_{nq} \rho_{qp} \left(e^{+\frac{ief}{\hbar c}}\right)_{pm} ,
\end{equation} 
where the scalar function $f(t,\mb{r})$ determines the gauge transformation of the potentials
\begin{align}
\label{Eq:gauge_transformation}
\tilde{\mb{A}} = \mb{A} + \nabla f(t,\mb{r}), 
\hspace{1cm} 
\tilde{\varphi} = \varphi - \frac{1}{c} \frac{\partial f(t,\mb{r})}{\partial t}. 
\end{align}
In particular, the transformation from the vector potential (\ref{Eq:vector_P}) to scalar potential (\ref{Eq:scalar_P}) is 
\begin{equation}
\label{Eq:f_Ap2phip}
\nabla f = -\mb{A}_P .
\end{equation}
Within the linear approximation with respect to $f$ we obtain from Eq.~(\ref{Eq:rho_gauge_transformation}):
\begin{equation}
\label{Eq:rho1_transfrom_linear}
\rho^{1,A_P}_{nm} \Rightarrow \rho^{1,A_P}_{nm} - \frac{i e}{\hbar c} f_{nm} (n_m - n_n) .
\end{equation}
Next, we will use the general relationship (see e.g. \cite{tokman2009})
\begin{equation}
f_{nm} = \frac{-i\hbar}{E_n - E_m} \left( \frac{\nabla f \cdot \hat{\mb{v}} + \hat{\mb{v}} \cdot \nabla f}{2} \right)_{nm} ,
\end{equation}
from which we obtain from $\hat{\mb{v}} = v_F \hat{\mb{\sigma}}$ that
\begin{equation}
\label{Eq:fnm_graphene}
f_{nm} = \frac{-i\hbar v_F \left(\hat{\mb{\sigma}} \cdot \nabla f\right)_{nm}}{E_n-E_m} .
\end{equation}
As a result, from Eqs.~(\ref{Eq:rho1_transfrom_linear}), (\ref{Eq:fnm_graphene}) and (\ref{Eq:f_Ap2phip}) one gets
\begin{equation}
\label{Eq:rho1_transfrom_Ap2phip}
\rho^{1,A_P}_{nm}(\omega) \Rightarrow \rho^{1,A_P}_{nm}(\omega) + \frac{e v_F}{c} \frac{ \left[\hat{\sigma}_x A_x(\omega)\right]_{nm} (n_m - n_n)}{E_n - E_m} .
\end{equation}
Taking into account Eq.~(\ref{Eq:rho_linear}), Eq.~(\ref{Eq:rho1_transfrom_Ap2phip}) can be represented as $\rho^{1,A_P}_{nm}(\omega) \Rightarrow \rho^{1,A_P}_{nm}(\omega) - \rho^{1,A_P}_{nm}(\omega \rightarrow 0)$, which is identical to Eq.~(\ref{Eq:jq_Falkovsky}).

The structure of transformation (\ref{Eq:rho_gauge_transformation}) makes it clear why the density matrix with an infinite trace can give rise to the divergent current. Consider the density matrix in the form $\rho_{nm} = \rho_{mm}\delta_{nm} + \xi_{n \neq m}$, where $\xi$ is a small perturbation. The sum $\sum_{mn} \mb{j}_{mn} \xi_{nm} $ can converge in a certain gauge even if the trace $\sum_m \rho_{mm} $ diverges. However, the transformation (\ref{Eq:rho_gauge_transformation}) to a different gauge projects the diagonal of the matrix with an infinite trace onto off-diagonal elements, which can lead to the divergence in Eq.~(\ref{Eq:sum_to_integral}). The inverse is also true: the divergence can be eliminated by the transformation (\ref{Eq:rho_gauge_transformation}) as we have just shown above. 

It is also clear that the separation of the response into intraband and interband components depends generally on the choice of the gauge since the transformation (\ref{Eq:rho_gauge_transformation}) mixes different contributions. At the same time, a correctly defined current has to be gauge-invariant.

\section{Second-order nonlinear response \label{Sec:2nd_order}}
Now we consider the second-order nonlinear response to the bichromatic field which we will represent through the vector potential in order to describe both P- and S-polarized fields with the same formalism. We will write the in-plane field components at frequencies $\omega_{1,2}$ directed along unit vectors $\bm{\eta}_{1,2}$ as 
\begin{align}
\mb{A} = \frac{1}{2} \mb{\eta}_1 A(\omega_1) e^{i(\mb{q}_1\cdot\mb{r}_{\|} - \omega_1 t)} +  \frac{1}{2} \mb{\eta}_2 A(\omega_2) e^{i(\mb{q}_2\cdot\mb{r}_{\|} - \omega_2 t)} + \mathrm{c.c.}
\end{align}
We need to calculate the perturbation of the density matrix at the sum frequency $\omega_1+\omega_2$. The term quadratic with respect to the field can be written as
\begin{widetext}
\begin{align}
&\phantom{{}={}}\rho_{mn}^{(2)}(\omega_1+\omega_2)  = \left(\frac{e}{2c}\right)
\frac{1}{\hbar(\omega_1+\omega_2)-(\epsilon_m-\epsilon_n)}   \nonumber \\
&\times \sum_{l\neq m,n} \left[ \left( (\hat{\mb{v}} \cdot \mb{\eta}_1) e^{i\mb{q}_1 \cdot \mb{r}} \right)_{ml} A(\omega_1) 
\rho_{ln}^{(1)}(\omega_2) - \rho_{ml}^{(1)}(\omega_1) \left( (\hat{\mb{v}} \cdot \mb{\eta}_2) e^{i\mb{q}_2 \cdot \mb{r}} \right)_{ln} A(\omega_2) \right]  + \left\{ 1 \leftrightarrow 2 \right\} \nonumber \\
&= 
\frac{1}{2} \left(\frac{e}{c}\right)^2 
\frac{ A(\omega_1) A(\omega_2) }{\hbar(\omega_1+\omega_2)-(\epsilon_m-\epsilon_n)} 
\times \sum_{l\neq m,n} \left( (\hat{\mb{v}} \cdot \mb{\eta}_1) e^{i\mb{q}_1 \cdot \mb{r}} \right)_{ml} 
\left( (\hat{\mb{v}} \cdot \mb{\eta}_2) e^{i\mb{q}_2 \cdot \mb{r}} \right)_{ln} \nonumber \\ 
& \times  \left[
\frac{ (\rho_{nn}-\rho_{ll}) }{ \hbar\omega_2 - (\epsilon_l - \epsilon_n) }
- \frac{ (\rho_{ll}-\rho_{mm}) }{  \hbar\omega_1 - (\epsilon_m - \epsilon_l) }
\right] 
+ \left\{ 1 \leftrightarrow 2 \right\} .
\end{align}
\end{widetext}

 The trace of the corresponding Fourier harmonic of the induced current can be then calculated as 
\begin{equation}
\label{Eq:jq_2nd}
\mb{j}^{(q_1+q_2)}(\omega_1+\omega_2) = - e \sum_{mn} \left(\hat{\mb{v}} e^{-i(\mb{q}_1+\mb{q}_2)\cdot\mb{r}}\right)_{nm} \rho_{mn}^{(2)}(\omega_1+\omega_2) .
\end{equation}
The second-order response at the difference frequency, $\rho^{(2)}_{mn}(\omega_1-\omega_2)$ can be obtained  by replacing
\begin{equation}
\omega_2 \Rightarrow -\omega_2, \; \bm{q}_2 \Rightarrow -\bm{q}_2, \; A(\omega_2) \Rightarrow A^*(\omega_2) .
\end{equation}

Next, we transform from summation to integration over $\bm{k}$-states, introduce the corresponding occupation numbers $f(s,\bm{k})$ of the momentum states in each band, apply the momentum conservation in a three-wave mixing process, and take into account spin and valley degeneracy. Note that the integral over the electron momenta converges, as opposed to the linear response calculations where one needs to regularize the integral by either subtracting the contribution at zero frequency or adding a $k^2$ term to the Hamiltonian, as discussed above. The result is 
\begin{align}
&\phantom{{}={}} \mb{J}^{(2)}(\omega_1+\omega_2)  \nonumber \\ 
&= - \frac{e^3 v_F^3}{16 \pi^2 c^2 \hbar^2}  A(\omega_1) A(\omega_2)  \sum_{s_m,s_n,s_l} \int d^2\mb{k} 
\frac{1}
{
(\omega_1+\omega_2)- v_F (s_m | \mb{k}+\mb{q}_1 |  - s_n |\mb{k}-\mb{q}_2| )
} \nonumber \\
&\times \left[
\frac{ f(s_n,|\mb{k}-\mb{q}_2|) - f(s_l,|\mb{k}|) }
{\omega_2 -  v_F (s_l |\mb{k}| - s_n |\mb{k}-\mb{q}_2|)} 
-
\frac{ f(s_l,|\mb{k}|) - f(s_m,|\mb{k}+\mb{q}_1|) }
{ \omega_1 -  v_F (s_m |\mb{k}+\mb{q}_1| - s_l |\mb{k}|) }
\right] \nonumber \\
&\times \left[ (\eta_{1x} - i\eta_{1y}) s_m e^{i\theta(\mb{k})} + (\eta_{1x} + i\eta_{1y}) s_l e^{-i\theta(\mb{k}+\mb{q}_1)} \right]  \nonumber \\
&\times \left[ (\eta_{2x} - i\eta_{2y}) s_l e^{i\theta(\mb{k}-\mb{q}_2)} + (\eta_{2x} + i\eta_{2y}) s_n e^{-i\theta(\mb{k})} \right] \nonumber \\
&\times \left[ (\mb{x}_0 + i \mb{y}_0) s_m e^{-i\theta(\mb{k}-\mb{q}_2)} + (\mb{x}_0 - i \mb{y}_0) s_n e^{i\theta(\mb{k}+\mb{q}_1)} \right]  \nonumber \\
&+ \left\{ 1 \leftrightarrow 2 \right\} .
\label{Eq:J_2ndorder_general} 
\end{align}

This equation can be integrated numerically for any given geometry of incident fields and electron distribution. We consider the limit of the Fermi distribution with a strong degeneracy, direct all in-plane photon wave vectors along x-axis,  and expand the integrand in Eq.~(\ref{Eq:J_2ndorder_general}) in powers of $q_1, q_2$. The integral over the term of zeroth-order in $q$ vanishes, as expected from symmetry. We will keep the terms linear in $q$.  Also we have to evaluate separately the intraband contribution $s_l=s_m=s_n$ and all types of mixed interband-intraband contributions:  $s_m = s_n = -s_l$, $s_m = s_l = -s_n$, and $s_n = s_l = -s_m$. 
After performing this procedure, we find the following nonzero components of the second-order nonlinear conductivity tensor, while all other components are zero: 
\begin{align}
&\phantom{{}={}}\sigma^{(2)}_{xxx}(\omega_1+\omega_2;\omega_1,\omega_2)  \nonumber \\
&= 
- s(\epsilon_F) \frac{e^3 v_F^2}{2 \pi \hbar^2}  \frac{1}{\omega_1^2 \omega_2^2 (\omega_1+\omega_2)} \frac{1}{(\omega_1^2 - 4 v_F^2 k_F^2) (\omega_2^2 - 4 v_F^2 k_F^2) ((\omega_1 + \omega_2)^2 - 4 v_F^2 k_F^2 )} \nonumber \\
&\times \left[ - 4 v_F^4 k_F^4 (q_1 \omega_2^3 (2 \omega_1 + \omega_2) + q_2 \omega_1^3 (\omega_1 + 2 \omega_2) ) + 16 v_F^6 k_F^6 (q_1 \omega_2 (2 \omega_1 + \omega_2) +  q_2 \omega_1 (\omega_1 + 2 \omega_2)) \right]
 ,     \label{xxx}    \\
%
&\phantom{{}={}}\sigma^{(2)}_{xyy}(\omega_1+\omega_2;\omega_1,\omega_2)  \nonumber \\
&= 
- s(\epsilon_F) \frac{e^3 v_F^2}{2 \pi \hbar^2}  \frac{1}{\omega_1^2 \omega_2^2 (\omega_1+\omega_2)} \frac{1}
{(\omega_1^2 - 4 v_F^2 k_F^2) (\omega_2^2 - 4 v_F^2 k_F^2) ((\omega_1 + \omega_2)^2 - 4 v_F^2 k_F^2 )} \nonumber \\
&\times \left[ 4(v_F k_F)^2 \omega_1 \omega_2 (\omega_1+\omega_2)^2 (q_1 \omega_2^2 + q_2 \omega_1^2) \right. \nonumber \\
&\phantom{{}={}} + 4 (v_F k_F)^4 (q_1 \omega_2^4 - (6 q_1 + 4 q_2) \omega_1 \omega_2^3 - 8 (q_1+q_2) \omega_1^2 \omega_2^2 - (4 q_1 + 6 q_2) \omega_1^3 \omega_2 + q_2 \omega_1^4 )  \nonumber \\
&\phantom{{}={}} + \left. 16 (v_F k_F)^6 ( q_1 \omega_2 ( 2 \omega_1 - \omega_2) + q_2 \omega_1 ( 2 \omega_2 -  \omega_1)) \right]  , \label{xyy}    \\
%
&\phantom{{}={}}\sigma^{(2)}_{yxy}(\omega_1+\omega_2;\omega_1,\omega_2)  \nonumber \\
&= 
- s(\epsilon_F) \frac{e^3 v_F^2}{2 \pi \hbar^2}  \frac{1}{\omega_1^2 \omega_2^2 (\omega_1+\omega_2)} \frac{1}{(\omega_1^2 - 4 v_F^2 k_F^2) (\omega_2^2 - 4 v_F^2 k_F^2) ((\omega_1 + \omega_2)^2 - 4 v_F^2 k_F^2 )} \nonumber \\
&\times \left[ 4(v_F k_F)^2 \omega_1^2 \omega_2 (\omega_1+\omega_2)( q_1 \omega_2^2 - q_2 \omega_1 (\omega_1 + 2 \omega_2)) \right. \nonumber \\
&\phantom{{}={}} + 4 (v_F k_F)^4 (q_2 \omega_1 (\omega_1 + 2 \omega_2)^3 - q_1 \omega_2 (4 \omega_1^3 + 4 \omega_1^2 \omega_2 + 2 \omega_1 \omega_2^2 + 3 \omega_2^3)) \nonumber \\
&\phantom{{}={}} + \left. 16 (v_F k_F)^6 ( q_1 \omega_2 (2 \omega_1 + 3 \omega_2) - q_2 \omega_1 (\omega_1+2 \omega_2) ) \right] , \label{yxy} \\
%
&\phantom{{}={}}\sigma^{(2)}_{yyx}(\omega_1+\omega_2;\omega_1,\omega_2)  \nonumber \\
&= 
- s(\epsilon_F) \frac{e^3 v_F^2}{2 \pi \hbar^2}  \frac{1}{\omega_1^2 \omega_2^2 (\omega_1+\omega_2)} \frac{1}{(\omega_1^2 - 4 v_F^2 k_F^2) (\omega_2^2 - 4 v_F^2 k_F^2) ((\omega_1 + \omega_2)^2 - 4 v_F^2 k_F^2 )} \nonumber \\
&\times \left[ 4(v_F k_F)^2 \omega_1 \omega_2^2 (\omega_1 + \omega_2) (q_2 \omega_1^2 - q_1 \omega_2 (2 \omega_1 + \omega_2)) \right. \nonumber \\
&\phantom{{}={}} + 4 (v_F k_F)^4 (q_1 \omega_2 (2 \omega_1 + \omega_2)^3 - q_2 \omega_1 (3 \omega_1^3 + 2 \omega_1^2 \omega_2 + 4 \omega_1 \omega_2^2 + 4 \omega_2^3)) \nonumber \\
&\phantom{{}={}} + \left. 16 (v_F k_F)^6 (  q_2 \omega_1 (3 \omega_1 + 2 \omega_2) - q_1 \omega_2 (2 \omega_1 + \omega_2) ) \right] . \label{yyx} 
\end{align}
Here $s(\epsilon_F) = \pm 1$ depending on whether the Fermi level is in the conduction or valence band.  A sketch of the second-order nonlinear process for an obliquely incident light of mixed polarization is shown in Fig.~1. Note that when both pump fields have either S- or P-polarization, the generated nonlinear current has only the $x$-component (along the in-plane direction of propagation of the pumps). When the polarizations are mixed, the $y$-component of the nonlinear current appears due to $yxy$ and $yyx$ components of the nonlinear conductivity (they are different only by permutation of indices 1 and 2 referring to the two pump fields). 

\begin{figure}[htb]
\begin{center}
\includegraphics[scale=0.7]{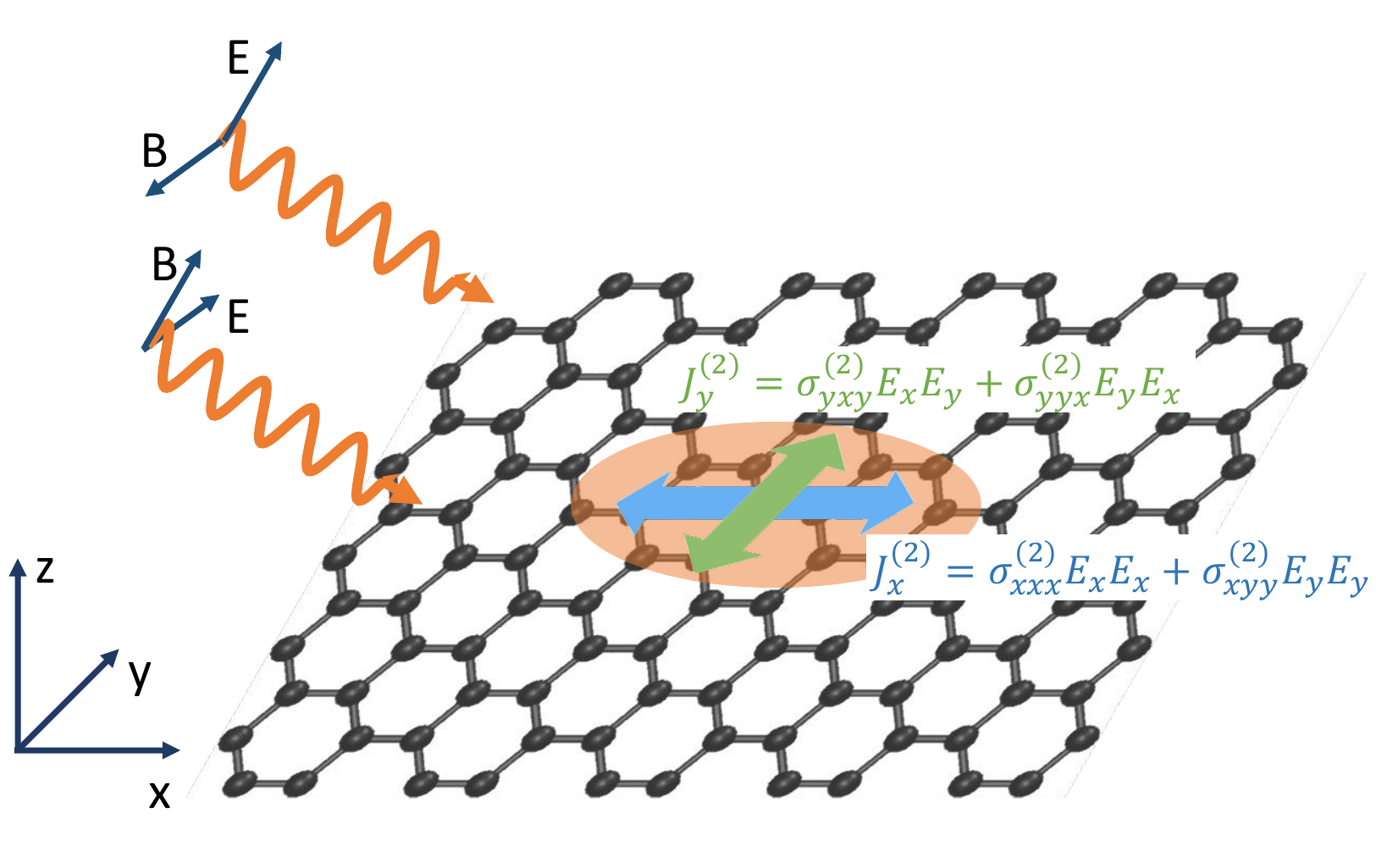}
\caption{A sketch of the second order nonlinear current generation in the graphene plane for obliquely incident light. }
\label{Fig:2nd_nonlinear_illustration}
\end{center}
\end{figure}

 Apparent ``non-reciprocity'' of the expressions for $\sigma^{(2)}_{yxx} = 0$ (P-in, S-out channel) and $\sigma^{(2)}_{xyy}$ (S-in, P-out channel) has a simple physical explanation: a P-polarized incident field cannot create a current orthogonal to the electric field, whereas an incident S-polarized field creates such a current via the magnetic field component $B_z$ normal to the layer.

The expressions for the nonlinear conductivity tensor that we obtained pass all symmetry and gauge invariance tests. Indeed, one can verify that the value of $\sigma^{(2)}_{xxx}$ agrees with the one derived using scalar potential in the interaction Hamiltonian. Furthermore, after converting the nonlinear conductivity to the nonlinear susceptibility according to  
$$ \chi^{(2)}_{ijk}(\omega_1+\omega_2;\omega_1,\omega_2) = \frac{i \sigma_{ijk}^{(2)}(\omega_1+\omega_2;\omega_1,\omega_2)}{\omega_1 + \omega_2 },
$$
one can verify that all components of the nonlinear susceptibility tensor satisfy proper permutation relations; see e.g. Ch.~2.9 in \cite{keldysh}:
\begin{align}
\label{permut}
\chi^{(2)}_{ijk}(\omega_3=\omega_1+\omega_2) = \chi^{(2)}_{jik}(-\omega_1=-\omega_3+\omega_2) = \chi^{(2)}_{kji}(-\omega_2=-\omega_3+\omega_1),
\end{align}
 where in-plane wave vectors have to be permuted together with frequencies.

The second-order response goes to zero when the Fermi energy $\epsilon_F$ goes to zero, and has maxima at resonances when one of 
the three frequencies involved in three-wave mixing is close to $2 \epsilon_F/\hbar = 2  v_F k_F$. Far from these resonances and for high frequencies or low doping, $2 v_F k_F \ll \omega_1, \omega_2, \omega_1+\omega_2$, expressions for the nonlinear conductivity are greatly simplified (we will give only the expressions for $\sigma^{(2)}_{xxx}$ and $\sigma^{(2)}_{xyy}$ for brevity):
\begin{align}
& \sigma^{(2)}_{xxx} = s(\epsilon_F) \frac{2e^3 v_F^2}{\pi \hbar^2} \frac{v_F^4 k_F^4 \left[ q_1 \omega_2^3 (2\omega_1+ \omega_2) + q_2 \omega_1^3 (2\omega_2 + \omega_1) \right] }{\omega_1^4 \omega_2^4 (\omega_1 + \omega_2)^3} ,
\label{high1} \\
& \sigma^{(2)}_{xyy} = - s(\epsilon_F) \frac{2e^3 v_F^2}{\pi \hbar^2} \frac{v_F^2 k_F^2 \left( q_1 \omega_2^2  + q_2 \omega_1^2\right) }{\omega_1^3 \omega_2^3 (\omega_1 + \omega_2)} . 
\label{high2}
\end{align}
An interesting and surprising result contained in these expressions is that the nonlinear frequency conversion of S-polarized
radiation into P-polarized radiation is much more efficient at high frequencies as compared to the P-in, P-out channel: $\sigma^{(2)}_{xyy}/\sigma^{(2)}_{xxx} \propto \frac{\omega^2}{v_F^2 k_F^2} \gg 1$. The dominance of the S-in, P-out channel is due to the chiral nature of Dirac fermion states. In particular, for the second-harmonic generation process $\omega_1 = \omega_2 = \omega$ and $q_1 = q_2 = q$, and the dominant component of the nonlinear conductivity tensor is simply 
\begin{align}
& \sigma^{(2)}_{xyy} = - s \frac{2e^3}{\pi \hbar^2} \frac{v_F^4 k_F^2 q}{\omega^5}.    
\label{shg}
\end{align}

In the opposite limit of low frequencies or high doping, $2 v_F k_F \gg \omega_1, \omega_2, \omega_1+\omega_2$, we also obtain simplified expressions: 
\begin{align}
& \sigma^{(2)}_{xxx} = s \frac{e^3 v_F^2}{4 \pi \hbar^2 \omega_1 \omega_2} \left( \frac{q_1 + q_2}{\omega_1+ \omega_2}  + \frac{q_1}{ \omega_1} + \frac{q_2}{ \omega_2}  \right)  ,
\label{low2-1} \\
& \sigma^{(2)}_{xyy} =  - s \frac{e^3 v_F^2}{4 \pi \hbar^2 \omega_1 \omega_2} \left[ \frac{q_1 + q_2}{\omega_1+ \omega_2}  + \frac{\omega_1 - \omega_2}{\omega_1 + \omega_2} \left(  \frac{q_1}{ \omega_1} + \frac{q_2}{ \omega_2} \right) \right]  .
\label{low2-2}
\end{align}
We verified that Eqs.~(\ref{low2-1}) and (\ref{low2-2}) can be derived independently from the single-band kinetic equation, i.e.~in the quasiclassical approximation described in Appendix D. This provides another test of our general expressions, since one should indeed expect that the single-band physics emerges in the limit of a strong doping and low frequencies, when all interband transitions become suppressed by Pauli blocking. Note that although Eqs.~(\ref{low2-1}) and (\ref{low2-2}) do not depend on $k_F$, they are valid only in the high-$k_F$ limit and are completely inapplicable for undoped graphene.  In fact, exact expressions (\ref{xxx}) and (\ref{xyy}) give $\sigma^{(2)}_{xxx} = 0$ and $\sigma^{(2)}_{xyy} = 0$  for $k_F = 0$, since in this case the nonlinear currents due to interband and intraband transitions cancel each other.  This can be viewed as a manifestation of the electron-hole symmetry in graphene. 

The nonlinear conductivity components (\ref{xxx}) and (\ref{xyy}) diverge when one or more of  
the three frequencies involved in three-wave mixing is close to $2 \epsilon_F/\hbar = 2  v_F k_F$. Close to resonance with $2 \epsilon_F/\hbar$ one has to include the imaginary part of the frequency which comes from the omitted relaxation term $-\gamma \rho_{mn}$ in the density-matrix equations. This amounts to substituting 
$\omega_1 \rightarrow \omega_1 + i \gamma_1$, $\omega_2 \rightarrow \omega_2 + i \gamma_2$, $\omega_1 + \omega_2 \rightarrow \omega_1 + \omega_2 + i \gamma_3$. Note that if we flip the sign of $\omega_2$ to describe the difference frequency generation process, the sign of $+ i \gamma_2$ remains the same, i.e. $\omega_2 \rightarrow -\omega_2 + i\gamma_2$. 

Even if dissipation is included, we can still use Eqs.~(\ref{permut}) to derive the  components of the nonlinear susceptibility tensor from other components. In order to do that, one needs to use Eqs.~(\ref{permut}) in the absence of dissipation and then add imaginary parts of frequencies. Of course the resulting expressions after adding dissipation won't satisfy the permutation relation Eqs.~(\ref{permut}).  

\begin{figure}[htb]
\begin{center}
\includegraphics[scale=0.7]{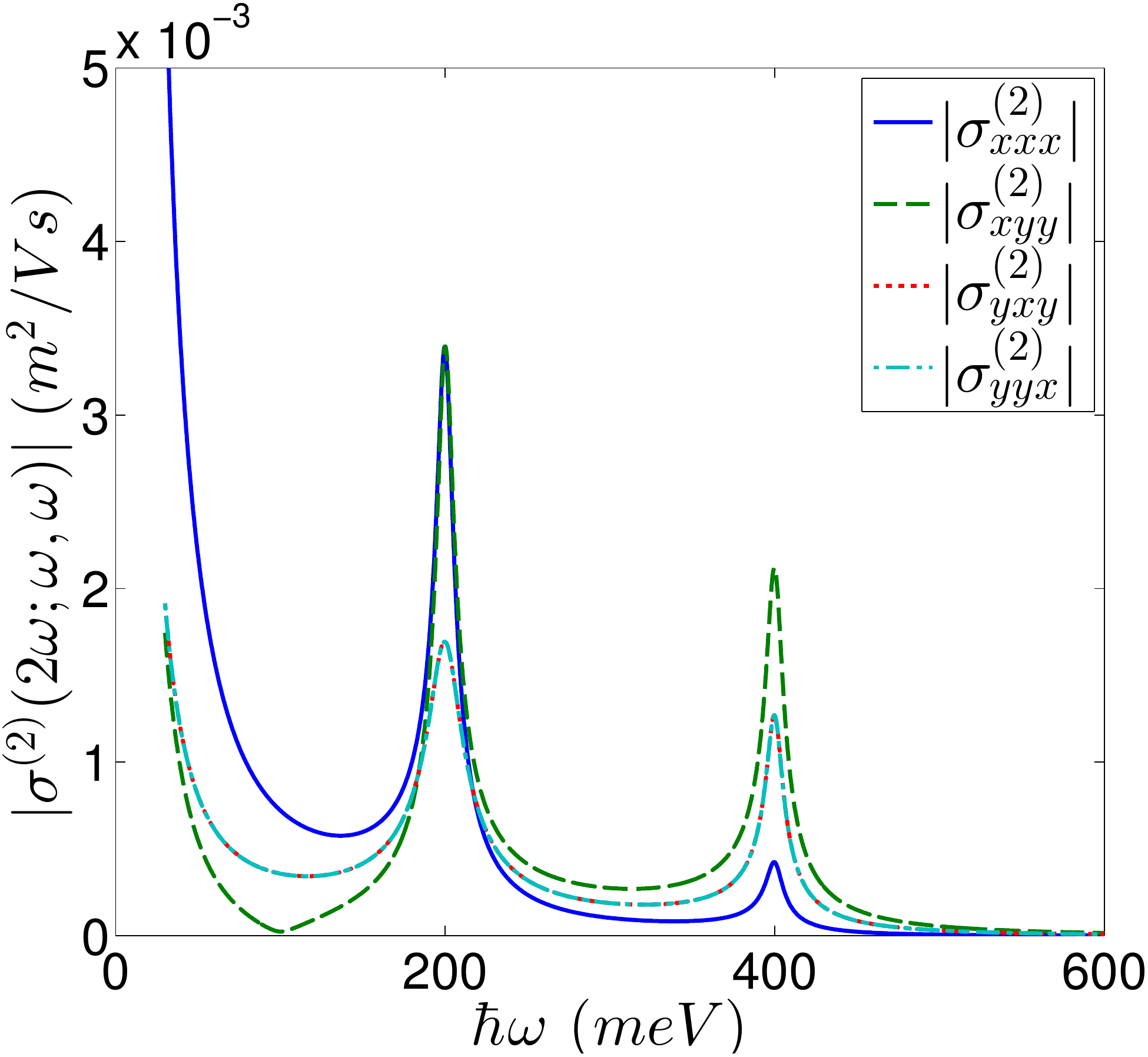}
\caption{Nonzero components of the second order nonlinear conductivity tensor for the process of SHG as a function of the fundamental frequency. The pump is incident at 45 degrees. The Fermi energy is 200 meV and all resonances are broadened by the same factor  $\gamma$ equal to 5 meV. }
\label{Fig:SHG}
\end{center}
\end{figure}
\begin{figure}[htb]
\begin{center}
\includegraphics[scale=0.7]{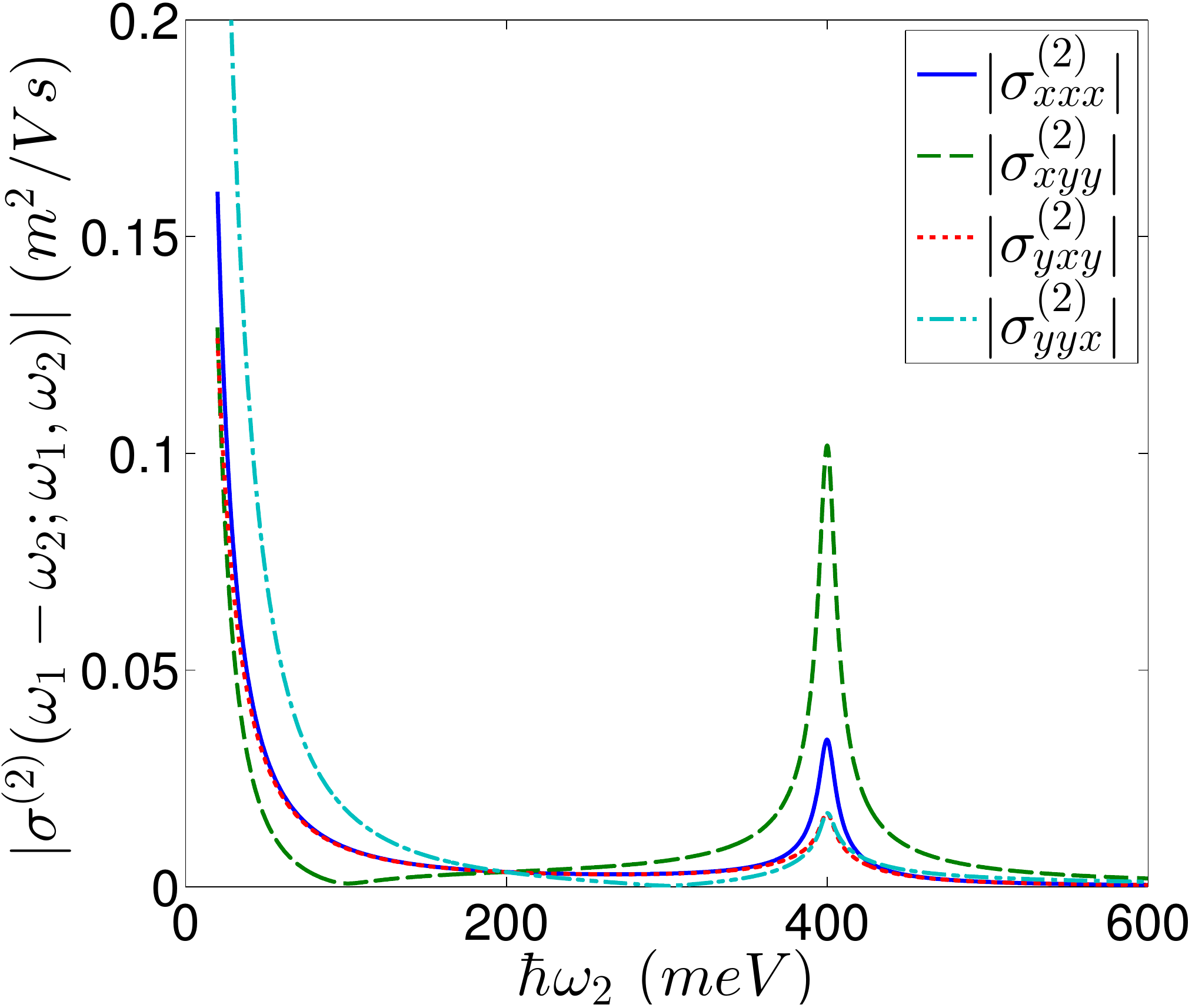}
\caption{Nonzero components of the second order nonlinear conductivity tensor for the process of DFG as a function of one of the pump frequencies ($\omega_2$). Frequency  $\omega_1$ is fixed at 400 meV.  Both pumps are incident in the $(xz)$-plane at 45 degrees. The Fermi energy is 200 meV and all resonances are broadened by the same factor  $\gamma$ equal to 5 meV.  }
\label{Fig:DFG}
\end{center}
\end{figure}
\begin{figure}[htb]
\begin{center}
\includegraphics[scale=0.7]{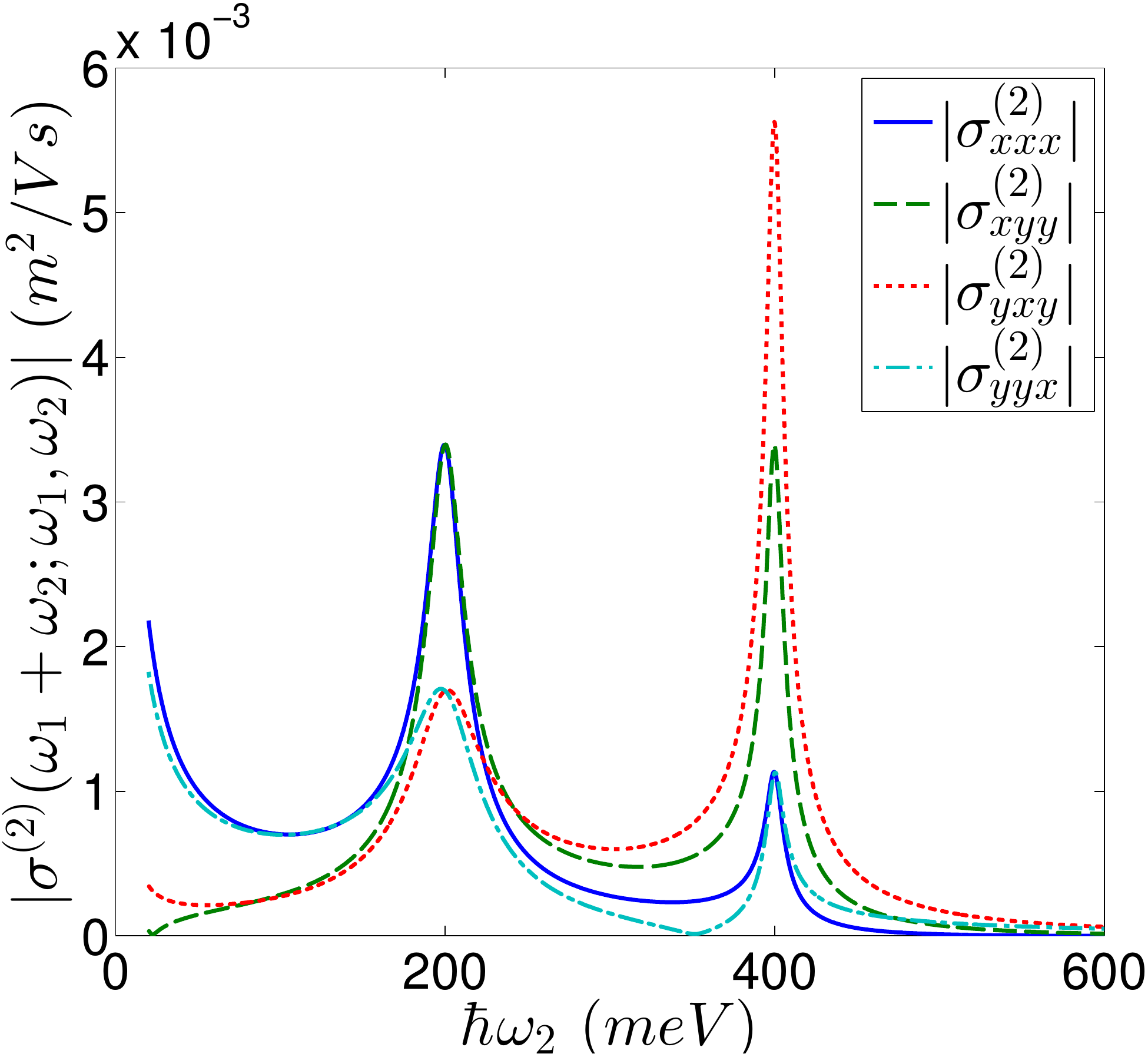}
\caption{Nonzero components of the second order nonlinear conductivity tensor for the process of SFG as a function of one of the pump frequencies ($\omega_2$). Frequency  $\omega_1$ is fixed at 200 meV.  Both pumps are incident in the $(xz)$-plane at 45 degrees. The Fermi energy is 200 meV and all resonances are broadened by the same factor  $\gamma$ equal to 5 meV.  }
\label{Fig:SFG}
\end{center}
\end{figure}

Figures 2-4 illustrate the above properties of the nonlinear conductivity for the processes of the second-harmonic generation (SHG), difference-frequency generation (DFG), and sum-frequency generation (SFG). We used SI units in the figures for easier comparison of the values with known materials. In Fig.~2, absolute values of nonzero   components of the nonlinear conductivity tensor for the SHG process $\omega + \omega \Rightarrow 2 \omega$ are plotted as a function of the fundamental frequency $\omega$, assuming that the Fermi energy is 200 meV and all resonances are broadened by the same half-width factor  $\gamma$ equal to 5 meV in energy units. The plots for $\sigma^{(2)}_{yxy}$ and $\sigma^{(2)}_{yyx}$ are identical as they should be. There are two prominent resonances at $\hbar \omega = 2 \epsilon_F = 400$ meV and $ 2\hbar  \omega = 2 \epsilon_F$. At high frequencies, the $xxx$ component falls off much faster than the $xyy$ component. At low frequencies, both components diverge as $1/\omega^2$. Our treatment, however, becomes invalid in the low-frequency limit $\omega \leq \gamma$ when any of the frequencies becomes lower than the scattering rate; that is why the plots are truncated at $\omega = 20$ meV. The quasi-classical method of the kinetic equation has the same applicability limit. 

Figure 3 shows absolute values of the nonzero components of the nonlinear conductivity tensor for the DFG process for the same values of $\epsilon_F$ and $\gamma$, as a function of $\omega_2$. The second frequency  $\hbar \omega_1$ is fixed to be 400 meV.  The same qualitative behavior is observed: there is a double resonance when both $\omega_1$ and $\omega_2$ are equal to $2 k_F v_F$. Note that there is no divergence at $\omega_1 - \omega_2 \rightarrow 0$ because the same factor $\omega_1 - \omega_2$ appears in the numerator.  There is divergence when $\omega_2 \rightarrow 0$ which should be truncated at $\omega_2 \sim \gamma$. 

In Fig.~4, the nonzero components of the nonlinear conductivity tensor for the SFG process are shown  as a function of $\omega_2$. The second frequency  $\hbar \omega_1$ is fixed at 200 meV.  As expected, all components show strong resonances when one of the frequencies or their sum is equal to $2 \epsilon_F = 400$ meV.

 The magnitude of the nonlinear response generally increases rapidly when one or more of the frequencies is decreased, as is obvious also from analytic expressions. For the DFG process, the magnitude of the nonlinear conductivity components is two orders of magnitude higher as compared to SHG or SFG.  As one of the frequencies goes to zero, the treatment becomes invalid, but one  could get an order of magnitude estimate of the maximum nonlinear conductivity by putting this frequency equal to $\gamma$. Using the same value of $\gamma = 5$ meV, one gets the nonlinear conductivity for DFG  of the order of several m$^2$/(Vs) in the THz range. This is a 2D conductivity. Purely for the sake of comparison with known bulk nonlinear materials, we can convert it to the bulk nonlinear susceptibility dividing by the frequency and the monolayer thickness of 0.3 nm, to arrive at $|\chi^{(2)}_{3D}| \sim 10^{-3}$ m/V. This is a huge value as compared to 1-100 pm/V values for most materials. Of course, only the 2D values of the graphene conductivity or susceptibility enter all physical results such as the intensity of the generated nonlinear signal \cite{yao2014,tokman2016} or the parametric gain \cite{tokman2016}. Still, combination of intrinsically large nonlinear conductivity of graphene and a surface plasmon resonance for the nonlinear signal may lead to quite significant efficiency of the nonlinear processes, as emphasized in the theoretical proposals \cite{yao2014,tokman2016}. 

In conclusion, we developed the full quantum-mechanical theory of the in-plane second-order nonlinear response of graphene beyond the electric dipole approximation. We provided a systematic derivation of the second-order nonlinear conductivity tensor, valid for all second-order processes, all frequencies and doping densities, as long as the massless Dirac fermion approximation for a single-particle Hamiltonian is applicable. Our approach can be applied to any system of massless chiral Dirac fermions, for example surface states in topological insulators such as Bi$_2$Se$_3$. We derived useful analytic expressions for the components of the nonlinear conductivity tensor, which satisfy all symmetry and permutation properties, and have a correct quasi-classical limit.  We also summarized main features of the linear response, with emphasis on its gauge properties and regularization.

\begin{acknowledgments} 
This work has been supported by the Air Force Office for Scientific Research through grants FA9550-15-1-0153 and  FA9550-14-1-0376. M.T.~acknowledges support from the Russian Foundation for Basic Research Grant No.~14-22-02034 and thanks I.D.~Tokman for helpful discussions. 
 \end{acknowledgments}

\appendix

\section{Evaluation of the linear current \label{appendix:linear}}
To calculate the current in the linear approximation with respect to the electromagnetic (EM) field, we will use 
Eqs.~(\ref{Eq:jq_matrix}), (\ref{Eq:eiqx_matrix}), (\ref{Eq:sigma_eiqx_matrix}), (\ref{Eq:rho_linear}), and (\ref{Eq:jq_linear}).  Assuming that the photon wave vector is much smaller than typical wave vectors of electrons, $q \ll k$, we calculate the following quantities in the zeroth and first order in $q$:
\begin{gather} 
n_{(\mb{k}+\mb{q})(s=+1)} - n_{\mb{k}(s=+1)} \approx q \cos\theta(\mb{k}) \frac{\partial n_{\mb{k}(+1)}}{\partial k} , \\
E_{(\mb{k}+\mb{q})(s=+1)} - E_{\mb{k}(s=+1)} \approx \hbar v_F q \cos\theta(\mb{k}) , \\
\frac{1}{2} \mb{j}^{(q)}_{\mb{k}(\mb{k}+\mb{q})(+1)(+1)} \approx - e v_F \left[ \mb{x}_0 \cos\theta(\mb{k}) + \mb{y}_0 \sin\theta(\mb{k}) \right] ,  \\
\frac{1}{2} \mb{j}^{(q)}_{\mb{k}(\mb{k}+\mb{q})(+1)(-1)} \approx - i e v_F \left[ \mb{x}_0 \sin\theta(\mb{k}) - \mb{y}_0 \cos\theta(\mb{k}) \right] .
\end{gather}
Consider first the EM field determined through a scalar potential. In this case we can  replace in Eq. (\ref{Eq:rho_linear})
\begin{equation}
\left[ \hat{V}(\omega) e^{i q x} \right]_{(\mb{k}+\mb{q})\mb{k}s s'} \approx - \frac{e\phi(\omega)}{4} \left[ i\frac{q}{k} \sin\theta(\mb{k}) + 1 + s s' \right] .
\end{equation} 
The summation in Eq.~(\ref{Eq:jq_linear}), can be replaced by integration using Eq.~(\ref{Eq:sum_to_integral}). Keeping the terms of the first order in $q$ in the conduction band, the integral can be transformed as $\int_0^\infty(\partial n_{k(+1)}/\partial k) k d k = - \int_0^\infty n_{k(+1)} d k = - k_F$. Introducing relaxation through the substitution $\omega \rightarrow \omega+i\gamma$, we arrive at Eqs.~(\ref{Eq:conductivity_linear_intra}) and (\ref{Eq:conductivity_linear_inter}).

Now we determine the EM field through the vector potential, in which case we should substitute the following in Eq.~(\ref{Eq:rho_linear}):
\begin{equation}
\left[ \hat{V}(\omega) e^{i q x} \right]_{(\mb{k}+\mb{q})\mb{k}s s'} \approx \frac{e v_F}{4 c} \left[ A_x \left( s e^{i\theta(\mb{k})} + s' e^{-i\theta(\mb{k})} \right) - i A_y \left( s e^{i\theta(\mb{k})} - s' e^{-i\theta(\mb{k})} \right) \right] .
\end{equation} 
After exactly the same steps as in the case of a scalar potential, we arrive at
\begin{gather}
\label{Eq:jq_intra}
\mb{j}^{(q)}_{(intra)} (\omega) = - \frac{g v_F^2 e^2 (\mb{x}_0 A_x + \mb{y}_0 A_y)}{4\pi^2\hbar c} \int_0^{2\pi} \frac{q \cos^3\theta d\theta}{\omega-v_F q \cos\theta} \int_0^{k_F} n_{k(+1)} d k , \\
\label{Eq:jq_inter}
\mb{j}^{(q)}_{(inter)} (\omega) = \frac{g v_F^2 e^2 (\mb{x}_0 A_x + \mb{y}_0 A_y)}{4\pi\hbar c} \int_0^\infty \left( \frac{1}{\omega+2 k v_F} - \frac{1}{\omega-2 k v_F} \right) \left( n_{k(-1)} - n_{k(+1)} \right) k d k .
\end{gather}
Note that $\mb{j}^{(q)}_{(inter)} (\omega) \rightarrow \infty$ when $\int_0^\infty n_{k(-1)} k dk \rightarrow \infty$. Therefore the current needs to be renormalized. Applying the renormalization Eq.~(\ref{Eq:jq_Falkovsky}), we obtain
\begin{align}
\mb{j}^{(q)}_{(intra)} (\omega) &= - \frac{g v_F e^2 \omega (\mb{x}_0 A_x + \mb{y}_0 A_y)}{4\pi^2\hbar c} \int_0^{2\pi} \frac{\cos^2\theta d\theta}{\omega-v_F q \cos\theta} \int_0^{k_F} n_{k(+1)} d k  \nonumber \\
& \approx
- \frac{g v_F e^2 (\mb{x}_0 A_x + \mb{y}_0 A_y)}{4\pi\hbar c} \int_0^{k_F} n_{k(+1)} d k  , \\
\mb{j}^{(q)}_{(inter)} (\omega) &= -\frac{g v_F e^2 \omega (\mb{x}_0 A_x + \mb{y}_0 A_y)}{8\pi\hbar c} \int_0^\infty \left( \frac{1}{\omega+2 k v_F} - \frac{1}{\omega-2 k v_F} \right) \left( n_{k(-1)} - n_{k(+1)} \right) d k ,
\end{align}
which again yields the expressions given in Sec.~\ref{Sec:linear}. 

If we choose the carrier distribution limited not only in the conduction band but also in the valence band, i.e. $n_{k(-1)} = 0$ for $k > k_{max;(-1)}$, then for the P-polarized field that can be defined through both scalar and vector potentials the sum $\mb{j}^{(q)}_{(intra;+1)} + \mb{j}^{(q)}_{(intra;(-1))} + \mb{j}^{(q)}_{(inter)}$ is invariant and finite without regularization with Eq.~(\ref{Eq:jq_Falkovsky}). This corroborates our conclusion that for massless Dirac fermions the need in renormalization (\ref{Eq:jq_Falkovsky}) is due to the bottomless valence band filled with electrons to infinite energies and wave vectors, which is an artifact of the model Hamiltonian (\ref{Eq:Hamiltonian}).

\section{How to correctly define current in a system with a massless Dirac spectrum \label{appendix:current_define}}

The prescription Eq.~(\ref{Eq:jq_Falkovsky}) for renormalization of the diverging linear current in a system of massless Dirac fermions can be justified if we consider a system with small deviation from the massless conical spectrum, for which the current becomes finite, and then let the deviation go to zero. Of course, the actual electron spectrum of graphene does deviate from the massless conical spectrum at high electron energies. However, it is reasonable to expect that at low enough energies any correction to the Hamiltonian (\ref{Eq:Hamiltonian}) becomes small, and all essential physics including the linear response is dominated by massless fermions. Therefore, it is important, at least from the methodological perspective, to provide physical justification of Eq.~(\ref{Eq:jq_Falkovsky}). 

Let's modify the Hamiltonian (\ref{Eq:Hamiltonian}) by adding a quadratic correction to the massless Dirac spectrum $E = s\hbar v_F k$:
\begin{equation}
\label{Eq:H_with_quadratic}
\hat{H}_0(\hat{\mb{p}}) = v_F \hat{\mb{\sigma}} \cdot \hat{\mb{p}} + \epsilon \frac{\hat{\mb{p}}^2}{2} \cdot \hat{1} .
\end{equation}
This Hamiltonian leads to the energy spectrum given by Eq.~(\ref{Eq:dispersion_with_quad}), whereas the eigenstates Eq.~(\ref{Eq:eigen_state}) remain the same. We will also assume that the change in the energy spectrum in the conduction band is insignificant, since 
\begin{equation}
\label{Eq:quad_condition}
\epsilon \hbar k_F \ll v_F .
\end{equation}
At the same time, the spectrum of Eq.~(\ref{Eq:H_with_quadratic}) creates a ``bottom'' of the valence band at $k = K$, where
\begin{equation}
\label{Eq:k_cutoff}
\epsilon \hbar K = v_F .
\end{equation}
Therefore, the integral over $k$-states in the valence band has now finite limits. 

In the presence of an EM field given by the vector potential  $\mb{A}$, one needs to replace $\hat{\mb{p}} \Rightarrow \hat{\mb{p}} + \frac{e}{c} \mb{A}$ in the Hamiltonian:
\begin{equation}
\label{Eq:H_with_quadratic_EM}
\hat{H}_0(\hat{\mb{p}}) = v_F \hat{\mb{\sigma}} \cdot \left( \hat{\mb{p}} + \frac{e}{c} \mb{A} \right) + \epsilon \frac{\left(\hat{\mb{p}} + \frac{e}{c} \mb{A}\right)^2}{2} \cdot \hat{1} .
\end{equation}
The resulting velocity operator, 
\begin{equation}
\hat{\mb{v}} = \frac{i}{\hbar} \left[ \hat{H}, \hat{\mb{r}} \right] = v_F \hat{\mb{\sigma}} + \epsilon \left(\hat{\mb{p}} + \frac{e}{c} \mb{A}\right) \cdot \hat{1} ,
\end{equation}
and the current operator,
\begin{equation}
\label{Eq:current_op_with_quad}
\hat{\mb{j}} = -e \hat{\mb{v}} = -e \left[ v_F \hat{\mb{\sigma}} + \epsilon \left(\hat{\mb{p}} + \frac{e}{c} \mb{A}\right) \cdot \hat{1} \right] 
\end{equation}
acquire a component which depends on the vector potential:
\begin{equation}
\delta \hat{\mb{j}} = -\epsilon \frac{e^2}{c} \mb{A} \cdot \hat{1} .
\end{equation}

Consider for definiteness an EM field given by the second of Eq.~(\ref{Eq:EM_potentials}) with $A_x = 0$, and also keep only the solution in zeroth order in $q/k$.

A new, $\mb{A}$-dependent component of the current operator $\delta \hat{\mb{j}}$ gives rise to an additional component of the linear current (see, e.g., \cite{tokman2009}):
\begin{equation}
\delta j_y = - \frac{\epsilon e^2 A_y e^{-i\omega t}}{2 c} \sum_{\mb{k}} n_{\mb{k}(s=-1)} + \mathrm{C.C.} ,
\end{equation}
where
\begin{equation}
\sum_{\mb{k}} n_{\mb{k}(s=-1)} = \frac{g}{4\pi^2} \int_0^{2\pi} d\theta \int_0^K n_{\mb{k}(-1)} k d k ,
\end{equation}
and the value of $K$ is determined by Eq.~(\ref{Eq:k_cutoff}). In the limit of Eq.~(\ref{Eq:quad_condition}) we can keep only the contribution of the valence band to the current component $\delta j_y$. This gives  (in the limit of strong degeneracy) 
\begin{equation}
\label{Eq:deltajy}
\delta j_y = - \frac{A_y e^{-i\omega t}}{2 c} \frac{g v_F e^2}{4\pi\hbar} \int_0^K n_{\mb{k}(-1)} d k  + \mathrm{C.C.} 
\end{equation}
Equation~(\ref{Eq:deltajy}) can be represented as a sum of two terms:
\begin{align}
&\phantom{{}={}} - \frac{g v_F e^2}{4\pi\hbar} \frac{A_y e^{-i\omega t}}{2 c} \int_0^K n_{\mb{k}(-1)} d k  \nonumber \\
&= - \frac{g v_F^2 e^2}{4\pi\hbar} \frac{A_y e^{-i\omega t}}{2 c} \int_0^K \left( \frac{1}{2 k v_F} - \frac{1}{-2 k v_F} \right) (n_{\mb{k}(-1)} - n_{\mb{k}(+1)}) k d k  \nonumber \\
&+ \frac{g v_F^2 e^2}{4\pi^2\hbar} \frac{A_y e^{-i\omega t}}{2 c} \int_0^{2\pi} \frac{q\cos^2\theta \cos\theta d\theta}{-v_F q \cos\theta} \int_0^{k_F} n_{\mb{k}(+1)} d k ,
\end{align}
where for a degenerate electron gas $n_{\mb{k}(+1)} = 0$ for $k > k_F$. Let us now compare this current component with the expressions (\ref{Eq:jq_intra}) and (\ref{Eq:jq_inter}) for the linear current that we derived in Appendix~\ref{appendix:linear} for a massless Dirac current ($\hat{\mb{j}} = - e v_F \hat{\mb{\sigma}}$), namely,  
\begin{align}
\label{Eq:jy_intra}
& j_y^{(intra)} = - \frac{g v_F^2 e^2}{4\pi^2\hbar} \frac{A_y e^{-i\omega t}}{2 c} \int_0^{2\pi} \frac{q\cos^2\theta \cos\theta d\theta}{\omega-v_F q \cos\theta} \int_0^{k_F} n_{\mb{k}(+1)} d k + \mathrm{C.C.} , \\
\label{Eq:jy_inter}
& j_y^{(inter)} = \frac{g v_F^2 e^2}{4\pi\hbar} \frac{A_y e^{-i\omega t}}{2 c} \int_0^K \left( \frac{1}{\omega + 2 k v_F} - \frac{1}{\omega - 2 k v_F} \right) (n_{\mb{k}(-1)} - n_{\mb{k}(+1)}) k d k + \mathrm{C.C.}  
\end{align}
From comparing (\ref{Eq:deltajy}) with (\ref{Eq:jy_intra}), (\ref{Eq:jy_inter}), it is obvious that $-\delta j_y = j_y^{(intra)}(\omega \rightarrow 0) + j_y^{(inter)}(\omega \rightarrow 0)$, i.e., adding this current component to the total current as $j_y^{(intra)} + j_y^{(inter)} + \delta j_y$ is completely equivalent to the renormalization given by Eq.~(\ref{Eq:jq_Falkovsky}) in the limit $K \rightarrow \infty$ which corresponds to the limit $\epsilon \rightarrow 0$. Note also that the current component $\hat{\mb{j}} = - \epsilon e  \hat{\mb{p}}$ which we neglected in Eq.~(\ref{Eq:current_op_with_quad}) becomes negligible as compared to $j_y^{(intra)} + j_y^{(inter)}$ in the same limit $\epsilon \rightarrow 0$. Actually this term vanishes since the distributions $n_{\mb{k}(-1)}$ and $n_{\mb{k}(+1)}$ don't depend on the direction of $\mb{k}$.

\section{Gauge transformation properties for massless Dirac systems \label{appendix:gauge_transform}}
We start from the Schr\"{o}dinger equation
\begin{equation}
\label{Eq:Sch_eq_gauge}
i\hbar \frac{\partial \mb{\Psi}}{\partial t} = \hat{H}(\mb{A},\varphi) \mb{\Psi}
\end{equation}
with the Hamiltonian of Eq.~(\ref{Eq:Hamiltonian_with_int}). Consider a gauge transformation of the field potentials from $(\mb{A},\varphi)$ to $(\tilde{\mb{A}},\tilde{\varphi})$. This transformation is determined by Eqs.~(\ref{Eq:gauge_transformation}) through a scalar function $f(\mb{r},t)$. Let $\tilde{\mb{\Psi}}$ be the solution of Eq.~(\ref{Eq:Sch_eq_gauge}) for $\hat{H}(\tilde{\mb{A}},\tilde{\varphi})$. One can see by direct substitution that the spinor $\mb{\Psi}$ is transformed is the same way as a scalar state function: $\tilde{\mb{\Psi}} = e^{-i\frac{e}{\hbar c} f} \mb{\Psi}$ \cite{landau2013quantum} (we consider a particle with negative charge $-e$). This transformation conserves the quantum-mechanical average current $\mb{j} = -e v_F \langle \mb{\Psi} | \hat{\mb{\sigma}} | \mb{\Psi} \rangle = -e v_F \langle \tilde{\mb{\Psi}} | \hat{\mb{\sigma}} | \tilde{\mb{\Psi}} \rangle$.

To obtain gauge transformation rules for the density matrix, it is convenient to use its coordinate representation as $\hat{\rho}(\mb{r},\mb{r}')$ \cite{tokman2009}. Following the standard procedure \cite{landau2013quantum}, we obtain 
\begin{equation}
\label{Eq:rho_rrp_rep}
\hat{\rho}(\mb{r},\mb{r}') = \sum_{mn} \rho_{mn} \left[ \mb{\Psi}_m(\mb{r}) \mb{\Psi}_n^*(\mb{r}') \right] ,
\end{equation}
where the expression $\left[ \mb{\Psi}_m(\mb{r}) \mb{\Psi}_n^*(\mb{r}') \right]$ is a  matrix formed by the elements of spinors $\mb{\Psi}_m(\mb{r})$ and $\mb{\Psi}_n^*(\mb{r}')$. Therefore, the operator $\hat{\rho}(\mb{r},\mb{r}')$ is a matrix with elements dependent on the pair of arguments $(\mb{r},\mb{r}')$:
\begin{equation}
\hat{\rho}(\mb{r},\mb{r}') = 
\begin{pmatrix}
\rho_{11}(\mb{r},\mb{r}') & \rho_{12}(\mb{r},\mb{r}') \\
\rho_{21}(\mb{r},\mb{r}') & \rho_{22}(\mb{r},\mb{r}') .
\end{pmatrix}
\end{equation}
The equation of motion for the operator $\hat{\rho}(\mb{r},\mb{r}')$ has a standard form, which follows directly from Eq.~(\ref{Eq:Sch_eq_gauge}):
\begin{equation}
\label{Eq:EoM_rho_rrprime}
i\hbar \frac{\partial \hat{\rho}(\mb{r},\mb{r}')}{\partial t} = \hat{H} \hat{\rho}(\mb{r},\mb{r}') - \hat{\rho}(\mb{r},\mb{r}') \overleftarrow{\hat{H}'} ,
\end{equation}
where the operator $\hat{H}$ acts only on the arguments $\mb{r}$ , whereas $\hat{H}'$ acts only on $\mb{r}'$, and the arrow above it means acting from right to left. The quantum-mechanical average of any operator $\hat{\mb{\Theta}}$ can be written in the matrix representation as $\mb{\Theta} = \sum_{mn} \mb{\Theta}_{nm}\rho_{mn}$, and in the coordinate representation as $\mb{\Theta} = \int_A d^2 \mb{r} \int_{A'} d^2\mb{r}' \left\{ \delta(\mb{r}-\mb{r}') \left[ \hat{\mb{\Theta}} \hat{\rho}(\mb{r},\mb{r}') \right] \right\}$, where it is assumed that the operator $\hat{\mb{\Theta}}$ acts only on $\mb{r}$.

Let $\tilde{\hat{\rho}}(\mb{r},\mb{r}')$ be the solution of Eq.~(\ref{Eq:EoM_rho_rrprime}) for the Hamiltonian $\hat{H}(\tilde{\mb{A}},\tilde{\varphi})$ given by Eq.~(\ref{Eq:Hamiltonian_with_int}). Then, following Ref.~\cite{tokman2009}, from Eq.~(\ref{Eq:EoM_rho_rrprime}) one can obtain  
\begin{equation}
\tilde{\hat{\rho}}(t,\mb{r},\mb{r}') = \hat{\rho}(t,\mb{r},\mb{r}') e^{-i u(t,\mb{r},\mb{r}')}, \hspace{1cm} u(t,\mb{r},\mb{r}') = \frac{e}{\hbar c} \left[ f(t,\mb{r}) - f(t,\mb{r}') \right] .
\end{equation}
Taking into account $\rho_{mn} = \langle \mb{\Psi}_m(\mb{r}) | \hat{\rho}(\mb{r},\mb{r}') | \mb{\Psi}_n(\mb{r}') \rangle $ which follows from Eq.~(\ref{Eq:rho_rrp_rep}), we arrive at Eq.~(\ref{Eq:rho_gauge_transformation}). 

Note that gauge transformation of the density matrix equation includes an appropriate transformation of the relaxation operator \cite{tokman2009}. The simplest approach which allows one to avoid complicated transformations is to neglect dissipation first, and then to replace $\omega \rightarrow \omega + i\gamma$ in the resulting expression for the dissipationless current. Of course, this approach works only for the simplest form of the relaxation operator in the relaxation time approximation.

\section{Quasiclassical approximation \label{appendix:quasiclassical}}
Here we provide the derivation of the quasiclassical equations of motion which allow one to derive Eqs. (\ref{low2-1},\ref{low2-2}) of the previous section in the single-band approximation of low frequencies and high Fermi energy, when the contribution of interband transitions can be neglected. 

In the absence of external fields ($\mb{A}=0$ ,$\varphi=0$) the solution of the Schr\"{o}dinger equation with Hamiltonian (\ref{Eq:Hamiltonian}) for a fixed energy of a quasiparticle can be written as
\begin{equation}
\mb{\Psi} = \begin{pmatrix} \Psi_1 \\ \Psi_2 \end{pmatrix} = \mathrm{const} \times \frac{e^{i\mb{k}\cdot\mb{r} - i \frac{E(\mb{k})}{\hbar}t}}{\sqrt{2}} \begin{pmatrix} s \\ e^{i\theta(\mb{k})} \end{pmatrix} ,
\label{Eq:sol_wo_field}
\end{equation}
where $s = \pm 1$, $E = s\hbar v_F |\mb{k}|$, $\theta(\mb{k})$ is an angle between the wave vector $\mb{k}$ and the x-axis. In the presence of the field, consider the solution of the Schr\"{o}dinger equation with Hamiltonian (\ref{Eq:Hamiltonian_with_int}) in the WKB approximation. Treating $\hbar$ as a small parameter, we seek the solution in the form close to (\ref{Eq:sol_wo_field}):
\begin{equation}
\mb{\Psi}(\mb{A},\varphi) = e^{\frac{i}{\hbar}S(t,\mb{r})} 
\left\{  
\begin{pmatrix} \Psi_1^{(0)}(t,\mb{r}) \\ \Psi_2^{(0)}(t,\mb{r}) \end{pmatrix} 
+
\hbar \begin{pmatrix} \Psi_1^{(1)}(t,\mb{r}) \\ \Psi_2^{(1)}(t,\mb{r}) \end{pmatrix} + \hbar^2 ...
\right\} .
\end{equation}
First consider the terms of zeroth order with respect to $\hbar$:
\begin{eqnarray}
\begin{array}{c}
(-\partial_t S + e\varphi) \Psi_1^{(0)} + v_F \left[ \left( -\partial_x S - \frac{e}{c} A_x \right) + i \left( \partial_y S + \frac{e}{c} A_y \right) \right] \Psi_2^{(0)} = 0,
\\
v_F \left[ \left( -\partial_x S - \frac{e}{c} A_x \right) - i \left( \partial_y S + \frac{e}{c} A_y \right) \right] \Psi_1^{(0)} + (-\partial_t S + e\varphi) \Psi_2^{(0)} = 0.
\end{array}
\label{Eq:SE_0th_order}
\end{eqnarray}
From (\ref{Eq:SE_0th_order}) we derive \\
({\bf {\lowercase\expandafter{\romannumeral 1 \relax}}}) the eikonal equation:
\begin{equation}
(-\partial_t S + e\varphi)^2 = v_F^2 \left( \frac{e}{c} \mb{A} + \mb{\nabla} S \right)^2 ,
\label{Eq:eikonal}
\end{equation}
({\bf {\lowercase\expandafter{\romannumeral 2 \relax}}}) the relationship between the spinor components:
\begin{equation}
\frac{\Psi_1^{(0)}}{\Psi_2^{(0)}} = \pm \left[ \cos\theta\left( \frac{e}{c} \mb{A} + \mb{\nabla} S \right) - i \sin\theta\left( \frac{e}{c} \mb{A} + \mb{\nabla} S \right) \right] = \pm e^{-i\theta\left( \frac{e}{c} \mb{A} + \mb{\nabla} S \right)} ,
\label{Eq:Psi12_ratio}
\end{equation}
where $\theta\left( \frac{e}{c} \mb{A} + \mb{\nabla} S \right)$ is the angle between vector $\frac{e}{c} \mb{A} + \mb{\nabla} S$ and the x-axis. Equation~(\ref{Eq:Psi12_ratio}) allows one to represent the WKB solution in the form
\begin{equation}
\mb{\Psi}(\mb{A},\varphi) = \frac{ \Phi(t,\mb{r}) e^{\frac{i}{\hbar}S(t,\mb{r})}}{\sqrt{2}} \begin{pmatrix}
s \\ e^{i\theta\left( \frac{e}{c} \mb{A} + \mb{\nabla} S \right)} .
\end{pmatrix}
\end{equation}
The expression for the factor $\Phi$ can be obtained by requiring that there exist the nontrivial solution to the next order term $\hbar \begin{pmatrix} \Psi_1^{(1)}(t,\mb{r}) \\ \Psi_2^{(1)}(t,\mb{r}) \end{pmatrix}$; this approach is used for example, in order to find the normal modes in anisotropic media \cite{ginzburg1970propagation}. However, it is much easier to use the conservation of the probability flux, which in our case is given by 
\begin{equation}
\frac{\partial}{\partial t} \left( \mb{\Psi}^* \mb{\Psi} \right) = - v_F \mb{\nabla} \cdot \left[ \mb{\Psi}^* \hat{\mb{\sigma}} \mb{\Psi} \right] .
\end{equation}
From here,
\begin{equation}
\frac{\partial |\Phi|^2}{\partial t} = - \mb{\nabla} \cdot \left[ \frac{s v_F \left( \frac{e}{c} \mb{A} + \mb{\nabla} S \right)}{\left| \frac{e}{c} \mb{A} + \mb{\nabla} S \right|} |\Phi|^2 \right] .
\end{equation}
Now consider the solution to the eikonal equation (\ref{Eq:eikonal}), which we will interpret as a Hamilton-Jacobi equation \cite{goldstein1980classical}, corresponding to the Hamiltonian $H(\mb{P},\mb{r},t)$:
\begin{align}
&\partial_t S + H(\mb{P},\mb{r},t) = 0, \hspace{1cm} \mb{\nabla} S = \mb{P}, \nonumber \\
&H(\mb{P},\mb{r},t) = -e\varphi(\mb{r},t) + s v_F \sqrt{\left(P_x + \frac{e}{c} A_x \right)^2 + \left(P_y + \frac{e}{c} A_y \right)^2 } .
\end{align}
The canonical equations of motion for this Hamiltonian are
\begin{eqnarray}
\begin{array}{c} \dot{\mb{r}} = \displaystyle  \frac{\partial H(\mb{P},\mb{r},t)}{\partial \mb{P}} = s v_F \frac{\mb{P} + \frac{e}{c}\mb{A}}{\left| \mb{P} + \frac{e}{c}\mb{A} \right|} ,  \\
\dot{\mb{P}} = \displaystyle - \frac{\partial H(\mb{P},\mb{r},t)}{\partial \mb{r}} = e \mb{\nabla} \varphi - \frac{e}{c} \left[ \dot{\mb{r}} \times (\mb{\nabla} \times \mb{A}) + (\dot{\mb{r}} \cdot \mb{\nabla} ) \mb{A} \right] .
\end{array}
\label{Eq:canonicalEoM}
\end{eqnarray}
Introducing the kinematic momentum $\mb{p} = \mb{P} + \frac{e}{c} \mb{A}$, for which $\dot{\mb{p}} = \dot{\mb{P}} + \frac{e}{c} \left[ \frac{\partial \mb{A}}{\partial t} + (\dot{\mb{r}} \cdot \mb{\nabla}) \mb{A} \right]$, we obtain from Eqs.~(\ref{Eq:canonicalEoM}) the quasiclassical equations of motion:
\begin{equation}
\dot{\mb{r}} = s v_F \frac{\mb{p}}{|\mb{p}|} ,  
\hspace{1cm} 
\dot{\mb{p}} = -e \mb{E} - \frac{e}{c} \left( \dot{\mb{r}} \times \mb{B} \right) ,
\label{Eq:quasiEoM}
\end{equation} 
where $\mb{E} = -\mb{\nabla} \varphi - \frac{1}{c} \frac{\partial \mb{A}}{\partial t}$, $\mb{B} = \mb{\nabla} \times \mb{A}$. Equations of motion (\ref{Eq:quasiEoM}) correspond to the kinetic equation for quasiparticles:
\begin{equation}
\frac{\partial f(\mb{r},\mb{p},t)}{\partial t} + s v_F \frac{\mb{p}}{|\mb{p}|} \frac{\partial f(\mb{r},\mb{p},t)}{\partial \mb{r}} - e \left[ \mb{E} + \frac{s v_F}{c} \left( \frac{\mb{p}}{|\mb{p}|}  \times \mb{B} \right) \right] \frac{\partial f(\mb{r},\mb{p},t)}{\partial \mb{p}} = \mathrm{St}[f(\mb{r},\mb{p},t)] ,
\label{Eq:kinetic_eq}
\end{equation}
where $\mathrm{St}[f(\mb{r},\mb{p},t)]$ is the collision integral. Quasiclassical equations (\ref{Eq:quasiEoM}) or (\ref{Eq:kinetic_eq}) were the starting point for evaluation of both linear and nonlinear optical response in a number of works; see, e.g. \cite{mikhailov2008,mikhailovSHG,glazov2011,smirnova2014,tokman2014}. As we have already discussed in the previous section, this approach can be justified only in the limit of low photon frequencies and large Fermi energies, when the contribution of interband transitions can be neglected.

\bibliography{graphene_2nd_order_nonlinear_Tokman_ref}

\begin{thebibliography}{26}%
\makeatletter
\providecommand \@ifxundefined [1]{%
 \@ifx{#1\undefined}
}%
\providecommand \@ifnum [1]{%
 \ifnum #1\expandafter \@firstoftwo
 \else \expandafter \@secondoftwo
 \fi
}%
\providecommand \@ifx [1]{%
 \ifx #1\expandafter \@firstoftwo
 \else \expandafter \@secondoftwo
 \fi
}%
\providecommand \natexlab [1]{#1}%
\providecommand \enquote  [1]{``#1''}%
\providecommand \bibnamefont  [1]{#1}%
\providecommand \bibfnamefont [1]{#1}%
\providecommand \citenamefont [1]{#1}%
\providecommand \href@noop [0]{\@secondoftwo}%
\providecommand \href [0]{\begingroup \@sanitize@url \@href}%
\providecommand \@href[1]{\@@startlink{#1}\@@href}%
\providecommand \@@href[1]{\endgroup#1\@@endlink}%
\providecommand \@sanitize@url [0]{\catcode `\\12\catcode `\$12\catcode
  `\&12\catcode `\#12\catcode `\^12\catcode `\_12\catcode `\%12\relax}%
\providecommand \@@startlink[1]{}%
\providecommand \@@endlink[0]{}%
\providecommand \url  [0]{\begingroup\@sanitize@url \@url }%
\providecommand \@url [1]{\endgroup\@href {#1}{\urlprefix }}%
\providecommand \urlprefix  [0]{URL }%
\providecommand \Eprint [0]{\href }%
\providecommand \doibase [0]{http://dx.doi.org/}%
\providecommand \selectlanguage [0]{\@gobble}%
\providecommand \bibinfo  [0]{\@secondoftwo}%
\providecommand \bibfield  [0]{\@secondoftwo}%
\providecommand \translation [1]{[#1]}%
\providecommand \BibitemOpen [0]{}%
\providecommand \bibitemStop [0]{}%
\providecommand \bibitemNoStop [0]{.\EOS\space}%
\providecommand \EOS [0]{\spacefactor3000\relax}%
\providecommand \BibitemShut  [1]{\csname bibitem#1\endcsname}%
\let\auto@bib@innerbib\@empty
\bibitem [{\citenamefont {Kumar}\ \emph {et~al.}(2013)\citenamefont {Kumar},
  \citenamefont {Kumar}, \citenamefont {Gerstenkorn}, \citenamefont {Wang},
  \citenamefont {Chiu}, \citenamefont {Smirl},\ and\ \citenamefont
  {Zhao}}]{kumar2013}%
  \BibitemOpen
  \bibfield  {author} {\bibinfo {author} {\bibfnamefont {N.}~\bibnamefont
  {Kumar}}, \bibinfo {author} {\bibfnamefont {J.}~\bibnamefont {Kumar}},
  \bibinfo {author} {\bibfnamefont {C.}~\bibnamefont {Gerstenkorn}}, \bibinfo
  {author} {\bibfnamefont {R.}~\bibnamefont {Wang}}, \bibinfo {author}
  {\bibfnamefont {H.-Y.}\ \bibnamefont {Chiu}}, \bibinfo {author}
  {\bibfnamefont {A.~L.}\ \bibnamefont {Smirl}}, \ and\ \bibinfo {author}
  {\bibfnamefont {H.}~\bibnamefont {Zhao}},\ }\href {\doibase
  10.1103/PhysRevB.87.121406} {\bibfield  {journal} {\bibinfo  {journal} {Phys.
  Rev. B}\ }\textbf {\bibinfo {volume} {87}},\ \bibinfo {pages} {121406}
  (\bibinfo {year} {2013})}\BibitemShut {NoStop}%
\bibitem [{\citenamefont {Hong}\ \emph {et~al.}(2013)\citenamefont {Hong},
  \citenamefont {Dadap}, \citenamefont {Petrone}, \citenamefont {Yeh},
  \citenamefont {Hone},\ and\ \citenamefont {Osgood}}]{hong2013}%
  \BibitemOpen
  \bibfield  {author} {\bibinfo {author} {\bibfnamefont {S.-Y.}\ \bibnamefont
  {Hong}}, \bibinfo {author} {\bibfnamefont {J.~I.}\ \bibnamefont {Dadap}},
  \bibinfo {author} {\bibfnamefont {N.}~\bibnamefont {Petrone}}, \bibinfo
  {author} {\bibfnamefont {P.-C.}\ \bibnamefont {Yeh}}, \bibinfo {author}
  {\bibfnamefont {J.}~\bibnamefont {Hone}}, \ and\ \bibinfo {author}
  {\bibfnamefont {R.~M.}\ \bibnamefont {Osgood}},\ }\href {\doibase
  10.1103/PhysRevX.3.021014} {\bibfield  {journal} {\bibinfo  {journal} {Phys.
  Rev. X}\ }\textbf {\bibinfo {volume} {3}},\ \bibinfo {pages} {021014}
  (\bibinfo {year} {2013})}\BibitemShut {NoStop}%
\bibitem [{\citenamefont {Hendry}\ \emph {et~al.}(2010)\citenamefont {Hendry},
  \citenamefont {Hale}, \citenamefont {Moger}, \citenamefont {Savchenko},\ and\
  \citenamefont {Mikhailov}}]{hendry2010}%
  \BibitemOpen
  \bibfield  {author} {\bibinfo {author} {\bibfnamefont {E.}~\bibnamefont
  {Hendry}}, \bibinfo {author} {\bibfnamefont {P.~J.}\ \bibnamefont {Hale}},
  \bibinfo {author} {\bibfnamefont {J.}~\bibnamefont {Moger}}, \bibinfo
  {author} {\bibfnamefont {A.~K.}\ \bibnamefont {Savchenko}}, \ and\ \bibinfo
  {author} {\bibfnamefont {S.~A.}\ \bibnamefont {Mikhailov}},\ }\href {\doibase
  10.1103/PhysRevLett.105.097401} {\bibfield  {journal} {\bibinfo  {journal}
  {Phys. Rev. Lett.}\ }\textbf {\bibinfo {volume} {105}},\ \bibinfo {pages}
  {097401} (\bibinfo {year} {2010})}\BibitemShut {NoStop}%
\bibitem [{\citenamefont {Gu}\ \emph {et~al.}(2012)\citenamefont {Gu},
  \citenamefont {Petrone}, \citenamefont {McMillan}, \citenamefont {van~der
  Zande}, \citenamefont {Yu}, \citenamefont {Lo}, \citenamefont {Kwong},
  \citenamefont {Hone},\ and\ \citenamefont {Wong}}]{gu2012}%
  \BibitemOpen
  \bibfield  {author} {\bibinfo {author} {\bibfnamefont {T.}~\bibnamefont
  {Gu}}, \bibinfo {author} {\bibfnamefont {N.}~\bibnamefont {Petrone}},
  \bibinfo {author} {\bibfnamefont {J.~F.}\ \bibnamefont {McMillan}}, \bibinfo
  {author} {\bibfnamefont {A.}~\bibnamefont {van~der Zande}}, \bibinfo {author}
  {\bibfnamefont {M.}~\bibnamefont {Yu}}, \bibinfo {author} {\bibfnamefont
  {G.-Q.}\ \bibnamefont {Lo}}, \bibinfo {author} {\bibfnamefont {D.-L.}\
  \bibnamefont {Kwong}}, \bibinfo {author} {\bibfnamefont {J.}~\bibnamefont
  {Hone}}, \ and\ \bibinfo {author} {\bibfnamefont {C.~W.}\ \bibnamefont
  {Wong}},\ }\href@noop {} {\bibfield  {journal} {\bibinfo  {journal} {Nature
  Photonics}\ }\textbf {\bibinfo {volume} {6}},\ \bibinfo {pages} {554}
  (\bibinfo {year} {2012})}\BibitemShut {NoStop}%
\bibitem [{\citenamefont {Sun}\ \emph {et~al.}(2010)\citenamefont {Sun},
  \citenamefont {Divin}, \citenamefont {Rioux}, \citenamefont {Sipe},
  \citenamefont {Berger}, \citenamefont {de~Heer}, \citenamefont {First},\ and\
  \citenamefont {Norris}}]{sun2010}%
  \BibitemOpen
  \bibfield  {author} {\bibinfo {author} {\bibfnamefont {D.}~\bibnamefont
  {Sun}}, \bibinfo {author} {\bibfnamefont {C.}~\bibnamefont {Divin}}, \bibinfo
  {author} {\bibfnamefont {J.}~\bibnamefont {Rioux}}, \bibinfo {author}
  {\bibfnamefont {J.~E.}\ \bibnamefont {Sipe}}, \bibinfo {author}
  {\bibfnamefont {C.}~\bibnamefont {Berger}}, \bibinfo {author} {\bibfnamefont
  {W.~A.}\ \bibnamefont {de~Heer}}, \bibinfo {author} {\bibfnamefont {P.~N.}\
  \bibnamefont {First}}, \ and\ \bibinfo {author} {\bibfnamefont {T.~B.}\
  \bibnamefont {Norris}},\ }\href {\doibase 10.1021/nl9040737} {\bibfield
  {journal} {\bibinfo  {journal} {Nano Letters}\ }\textbf {\bibinfo {volume}
  {10}},\ \bibinfo {pages} {1293} (\bibinfo {year} {2010})}\BibitemShut
  {NoStop}%
\bibitem [{\citenamefont {Glazov}\ and\ \citenamefont
  {Ganichev}(2014)}]{glazov2014}%
  \BibitemOpen
  \bibfield  {author} {\bibinfo {author} {\bibfnamefont {M.}~\bibnamefont
  {Glazov}}\ and\ \bibinfo {author} {\bibfnamefont {S.}~\bibnamefont
  {Ganichev}},\ }\href {\doibase
  http://dx.doi.org/10.1016/j.physrep.2013.10.003} {\bibfield  {journal}
  {\bibinfo  {journal} {Physics Reports}\ }\textbf {\bibinfo {volume} {535}},\
  \bibinfo {pages} {101} (\bibinfo {year} {2014})}\BibitemShut {NoStop}%
\bibitem [{\citenamefont {Bykov}\ \emph {et~al.}(2012)\citenamefont {Bykov},
  \citenamefont {Murzina}, \citenamefont {Rybin},\ and\ \citenamefont
  {Obraztsova}}]{bykov2012}%
  \BibitemOpen
  \bibfield  {author} {\bibinfo {author} {\bibfnamefont {A.~Y.}\ \bibnamefont
  {Bykov}}, \bibinfo {author} {\bibfnamefont {T.~V.}\ \bibnamefont {Murzina}},
  \bibinfo {author} {\bibfnamefont {M.~G.}\ \bibnamefont {Rybin}}, \ and\
  \bibinfo {author} {\bibfnamefont {E.~D.}\ \bibnamefont {Obraztsova}},\ }\href
  {\doibase 10.1103/PhysRevB.85.121413} {\bibfield  {journal} {\bibinfo
  {journal} {Phys. Rev. B}\ }\textbf {\bibinfo {volume} {85}},\ \bibinfo
  {pages} {121413} (\bibinfo {year} {2012})}\BibitemShut {NoStop}%
\bibitem [{\citenamefont {Cheng}\ \emph {et~al.}(2014)\citenamefont {Cheng},
  \citenamefont {Vermeulen},\ and\ \citenamefont {Sipe}}]{cheng2014}%
  \BibitemOpen
  \bibfield  {author} {\bibinfo {author} {\bibfnamefont {J.~L.}\ \bibnamefont
  {Cheng}}, \bibinfo {author} {\bibfnamefont {N.}~\bibnamefont {Vermeulen}}, \
  and\ \bibinfo {author} {\bibfnamefont {J.~E.}\ \bibnamefont {Sipe}},\ }\href
  {\doibase 10.1364/OE.22.015868} {\bibfield  {journal} {\bibinfo  {journal}
  {Opt. Express}\ }\textbf {\bibinfo {volume} {22}},\ \bibinfo {pages} {15868}
  (\bibinfo {year} {2014})}\BibitemShut {NoStop}%
\bibitem [{\citenamefont {Dean}\ and\ \citenamefont {van
  Driel}(2009)}]{dean2009}%
  \BibitemOpen
  \bibfield  {author} {\bibinfo {author} {\bibfnamefont {J.~J.}\ \bibnamefont
  {Dean}}\ and\ \bibinfo {author} {\bibfnamefont {H.~M.}\ \bibnamefont {van
  Driel}},\ }\href
  {http://scitation.aip.org/content/aip/journal/apl/95/26/10.1063/1.3275740}
  {\bibfield  {journal} {\bibinfo  {journal} {Applied Physics Letters}\
  }\textbf {\bibinfo {volume} {95}},\ \bibinfo {eid} {261910} (\bibinfo {year}
  {2009})}\BibitemShut {NoStop}%
\bibitem [{\citenamefont {Dean}\ and\ \citenamefont {van
  Driel}(2010)}]{dean2010}%
  \BibitemOpen
  \bibfield  {author} {\bibinfo {author} {\bibfnamefont {J.~J.}\ \bibnamefont
  {Dean}}\ and\ \bibinfo {author} {\bibfnamefont {H.~M.}\ \bibnamefont {van
  Driel}},\ }\href {\doibase 10.1103/PhysRevB.82.125411} {\bibfield  {journal}
  {\bibinfo  {journal} {Phys. Rev. B}\ }\textbf {\bibinfo {volume} {82}},\
  \bibinfo {pages} {125411} (\bibinfo {year} {2010})}\BibitemShut {NoStop}%
\bibitem [{\citenamefont {Mikhailov}(2011)}]{mikhailovSHG}%
  \BibitemOpen
  \bibfield  {author} {\bibinfo {author} {\bibfnamefont {S.~A.}\ \bibnamefont
  {Mikhailov}},\ }\href {\doibase 10.1103/PhysRevB.84.045432} {\bibfield
  {journal} {\bibinfo  {journal} {Phys. Rev. B}\ }\textbf {\bibinfo {volume}
  {84}},\ \bibinfo {pages} {045432} (\bibinfo {year} {2011})}\BibitemShut
  {NoStop}%
\bibitem [{\citenamefont {Glazov}(2011)}]{glazov2011}%
  \BibitemOpen
  \bibfield  {author} {\bibinfo {author} {\bibfnamefont {M.~M.}\ \bibnamefont
  {Glazov}},\ }\href {\doibase 10.1134/S0021364011070046} {\bibfield  {journal}
  {\bibinfo  {journal} {JETP Letters}\ }\textbf {\bibinfo {volume} {93}},\
  \bibinfo {pages} {366} (\bibinfo {year} {2011})}\BibitemShut {NoStop}%
\bibitem [{\citenamefont {Smirnova}\ \emph {et~al.}(2014)\citenamefont
  {Smirnova}, \citenamefont {Shadrivov}, \citenamefont {Miroshnichenko},
  \citenamefont {Smirnov},\ and\ \citenamefont {Kivshar}}]{smirnova2014}%
  \BibitemOpen
  \bibfield  {author} {\bibinfo {author} {\bibfnamefont {D.~A.}\ \bibnamefont
  {Smirnova}}, \bibinfo {author} {\bibfnamefont {I.~V.}\ \bibnamefont
  {Shadrivov}}, \bibinfo {author} {\bibfnamefont {A.~E.}\ \bibnamefont
  {Miroshnichenko}}, \bibinfo {author} {\bibfnamefont {A.~I.}\ \bibnamefont
  {Smirnov}}, \ and\ \bibinfo {author} {\bibfnamefont {Y.~S.}\ \bibnamefont
  {Kivshar}},\ }\href {\doibase 10.1103/PhysRevB.90.035412} {\bibfield
  {journal} {\bibinfo  {journal} {Phys. Rev. B}\ }\textbf {\bibinfo {volume}
  {90}},\ \bibinfo {pages} {035412} (\bibinfo {year} {2014})}\BibitemShut
  {NoStop}%
\bibitem [{\citenamefont {Yao}\ \emph {et~al.}(2014)\citenamefont {Yao},
  \citenamefont {Tokman},\ and\ \citenamefont {Belyanin}}]{yao2014}%
  \BibitemOpen
  \bibfield  {author} {\bibinfo {author} {\bibfnamefont {X.}~\bibnamefont
  {Yao}}, \bibinfo {author} {\bibfnamefont {M.}~\bibnamefont {Tokman}}, \ and\
  \bibinfo {author} {\bibfnamefont {A.}~\bibnamefont {Belyanin}},\ }\href
  {\doibase 10.1103/PhysRevLett.112.055501} {\bibfield  {journal} {\bibinfo
  {journal} {Phys. Rev. Lett.}\ }\textbf {\bibinfo {volume} {112}},\ \bibinfo
  {pages} {055501} (\bibinfo {year} {2014})}\BibitemShut {NoStop}%
\bibitem [{\citenamefont {Tokman}\ \emph {et~al.}(2016)\citenamefont {Tokman},
  \citenamefont {Wang}, \citenamefont {Oladyshkin}, \citenamefont {Kutayiah},\
  and\ \citenamefont {Belyanin}}]{tokman2016}%
  \BibitemOpen
  \bibfield  {author} {\bibinfo {author} {\bibfnamefont {M.}~\bibnamefont
  {Tokman}}, \bibinfo {author} {\bibfnamefont {Y.}~\bibnamefont {Wang}},
  \bibinfo {author} {\bibfnamefont {I.}~\bibnamefont {Oladyshkin}}, \bibinfo
  {author} {\bibfnamefont {A.~R.}\ \bibnamefont {Kutayiah}}, \ and\ \bibinfo
  {author} {\bibfnamefont {A.}~\bibnamefont {Belyanin}},\ }\href {\doibase
  10.1103/PhysRevB.93.235422} {\bibfield  {journal} {\bibinfo  {journal} {Phys.
  Rev. B}\ }\textbf {\bibinfo {volume} {93}},\ \bibinfo {pages} {235422}
  (\bibinfo {year} {2016})}\BibitemShut {NoStop}%
\bibitem [{\citenamefont {Constant}\ \emph {et~al.}(2016)\citenamefont
  {Constant}, \citenamefont {Hornett}, \citenamefont {Chang},\ and\
  \citenamefont {Hendry}}]{constant2016}%
  \BibitemOpen
  \bibfield  {author} {\bibinfo {author} {\bibfnamefont {T.~J.}\ \bibnamefont
  {Constant}}, \bibinfo {author} {\bibfnamefont {S.~M.}\ \bibnamefont
  {Hornett}}, \bibinfo {author} {\bibfnamefont {D.~E.}\ \bibnamefont {Chang}},
  \ and\ \bibinfo {author} {\bibfnamefont {E.}~\bibnamefont {Hendry}},\
  }\href@noop {} {\bibfield  {journal} {\bibinfo  {journal} {Nature Physics}\
  }\textbf {\bibinfo {volume} {12}},\ \bibinfo {pages} {124} (\bibinfo {year}
  {2016})}\BibitemShut {NoStop}%
\bibitem [{\citenamefont {Landau}\ and\ \citenamefont
  {Lifshitz}(2013)}]{landau2013quantum}%
  \BibitemOpen
  \bibfield  {author} {\bibinfo {author} {\bibfnamefont {L.~D.}\ \bibnamefont
  {Landau}}\ and\ \bibinfo {author} {\bibfnamefont {E.~M.}\ \bibnamefont
  {Lifshitz}},\ }\href {https://books.google.com/books?id=neBbAwAAQBAJ} {\emph
  {\bibinfo {title} {Quantum Mechanics: Non-Relativistic Theory}}},\
  Teoreticheskai︠a︡ fizika\ (\bibinfo  {publisher} {Elsevier Science},\
  \bibinfo {year} {2013})\BibitemShut {NoStop}%
\bibitem [{\citenamefont {Gantmakher}\ and\ \citenamefont
  {Levinson}(1987)}]{gantmakher1987}%
  \BibitemOpen
  \bibfield  {author} {\bibinfo {author} {\bibfnamefont {V.~F.}\ \bibnamefont
  {Gantmakher}}\ and\ \bibinfo {author} {\bibfnamefont {I.~B.}\ \bibnamefont
  {Levinson}},\ }\href {https://books.google.com/books?id=3MJ4AAAAIAAJ} {\emph
  {\bibinfo {title} {Carrier scattering in metals and semiconductors}}}\
  (\bibinfo  {publisher} {North-Holland, Amsterdam},\ \bibinfo {year}
  {1987})\BibitemShut {NoStop}%
\bibitem [{\citenamefont {Nair}\ \emph {et~al.}(2008)\citenamefont {Nair},
  \citenamefont {Blake}, \citenamefont {Grigorenko}, \citenamefont {Novoselov},
  \citenamefont {Booth}, \citenamefont {Stauber}, \citenamefont {Peres},\ and\
  \citenamefont {Geim}}]{nair2008}%
  \BibitemOpen
  \bibfield  {author} {\bibinfo {author} {\bibfnamefont {R.~R.}\ \bibnamefont
  {Nair}}, \bibinfo {author} {\bibfnamefont {P.}~\bibnamefont {Blake}},
  \bibinfo {author} {\bibfnamefont {A.~N.}\ \bibnamefont {Grigorenko}},
  \bibinfo {author} {\bibfnamefont {K.~S.}\ \bibnamefont {Novoselov}}, \bibinfo
  {author} {\bibfnamefont {T.~J.}\ \bibnamefont {Booth}}, \bibinfo {author}
  {\bibfnamefont {T.}~\bibnamefont {Stauber}}, \bibinfo {author} {\bibfnamefont
  {N.~M.~R.}\ \bibnamefont {Peres}}, \ and\ \bibinfo {author} {\bibfnamefont
  {A.~K.}\ \bibnamefont {Geim}},\ }\href {\doibase 10.1126/science.1156965}
  {\bibfield  {journal} {\bibinfo  {journal} {Science}\ }\textbf {\bibinfo
  {volume} {320}},\ \bibinfo {pages} {1308} (\bibinfo {year}
  {2008})}\BibitemShut {NoStop}%
\bibitem [{\citenamefont {Falkovsky}\ and\ \citenamefont
  {Varlamov}(2007)}]{falkovsky2007}%
  \BibitemOpen
  \bibfield  {author} {\bibinfo {author} {\bibfnamefont {A.~L.}\ \bibnamefont
  {Falkovsky}}\ and\ \bibinfo {author} {\bibfnamefont {A.~A.}\ \bibnamefont
  {Varlamov}},\ }\href {\doibase 10.1140/epjb/e2007-00142-3} {\bibfield
  {journal} {\bibinfo  {journal} {The European Physical Journal B}\ }\textbf
  {\bibinfo {volume} {56}},\ \bibinfo {pages} {281} (\bibinfo {year}
  {2007})}\BibitemShut {NoStop}%
\bibitem [{\citenamefont {Tokman}(2009)}]{tokman2009}%
  \BibitemOpen
  \bibfield  {author} {\bibinfo {author} {\bibfnamefont {M.~D.}\ \bibnamefont
  {Tokman}},\ }\href {\doibase 10.1103/PhysRevA.79.053415} {\bibfield
  {journal} {\bibinfo  {journal} {Phys. Rev. A}\ }\textbf {\bibinfo {volume}
  {79}},\ \bibinfo {pages} {053415} (\bibinfo {year} {2009})}\BibitemShut
  {NoStop}%
\bibitem [{\citenamefont {Il'inskii}\ and\ \citenamefont
  {Keldysh}(1994)}]{keldysh}%
  \BibitemOpen
  \bibfield  {author} {\bibinfo {author} {\bibfnamefont {Y.~A.}\ \bibnamefont
  {Il'inskii}}\ and\ \bibinfo {author} {\bibfnamefont {L.~V.}\ \bibnamefont
  {Keldysh}},\ }\href {https://books.google.com/books?id=XUr4VS91W60C} {\emph
  {\bibinfo {title} {Electromagnetic Response of Material Media}}}\ (\bibinfo
  {publisher} {Springer US},\ \bibinfo {year} {1994})\BibitemShut {NoStop}%
\bibitem [{\citenamefont {Ginzburg}(1970)}]{ginzburg1970propagation}%
  \BibitemOpen
  \bibfield  {author} {\bibinfo {author} {\bibfnamefont {V.~L.}\ \bibnamefont
  {Ginzburg}},\ }\href {https://books.google.com/books?id=vF55AAAAIAAJ} {\emph
  {\bibinfo {title} {The propagation of electromagnetic waves in plasmas}}},\
  International series of monographs on electromagnetic waves\ (\bibinfo
  {publisher} {Pergamon Press},\ \bibinfo {year} {1970})\BibitemShut {NoStop}%
\bibitem [{\citenamefont {Goldstein}(1980)}]{goldstein1980classical}%
  \BibitemOpen
  \bibfield  {author} {\bibinfo {author} {\bibfnamefont {H.}~\bibnamefont
  {Goldstein}},\ }\href {https://books.google.com/books?id=I1JKjwEACAAJ} {\emph
  {\bibinfo {title} {Classical mechanics}}},\ Addison-Wesley series in physics\
  (\bibinfo  {publisher} {Addison-Wesley},\ \bibinfo {year} {1980})\BibitemShut
  {NoStop}%
\bibitem [{\citenamefont {Mikhailov}\ and\ \citenamefont
  {Ziegler}(2008)}]{mikhailov2008}%
  \BibitemOpen
  \bibfield  {author} {\bibinfo {author} {\bibfnamefont {S.~A.}\ \bibnamefont
  {Mikhailov}}\ and\ \bibinfo {author} {\bibfnamefont {K.}~\bibnamefont
  {Ziegler}},\ }\href {http://stacks.iop.org/0953-8984/20/i=38/a=384204}
  {\bibfield  {journal} {\bibinfo  {journal} {Journal of Physics: Condensed
  Matter}\ }\textbf {\bibinfo {volume} {20}},\ \bibinfo {pages} {384204}
  (\bibinfo {year} {2008})}\BibitemShut {NoStop}%
\bibitem [{\citenamefont {Tokman}\ \emph {et~al.}(2014)\citenamefont {Tokman},
  \citenamefont {Erukhimova},\ and\ \citenamefont {Belyanin}}]{tokman2014}%
  \BibitemOpen
  \bibfield  {author} {\bibinfo {author} {\bibfnamefont {M.~D.}\ \bibnamefont
  {Tokman}}, \bibinfo {author} {\bibfnamefont {M.~A.}\ \bibnamefont
  {Erukhimova}}, \ and\ \bibinfo {author} {\bibfnamefont {A.}~\bibnamefont
  {Belyanin}},\ }\href {\doibase 10.1134/S0021364014180118} {\bibfield
  {journal} {\bibinfo  {journal} {JETP Letters}\ }\textbf {\bibinfo {volume}
  {100}},\ \bibinfo {pages} {390} (\bibinfo {year} {2014})}\BibitemShut
  {NoStop}%
\end{thebibliography}%

\end{document}